\documentclass{aa}  

\usepackage{graphicx}
\usepackage{txfonts}
\usepackage{hyperref}
\usepackage{orcidlink}
\makeatletter
\renewcommand*\aa@pageof{, page \thepage{} of \pageref*{LastPage}}
\makeatother

\begin{document}

   \title{Hidden shock powering the peak of SN~2020faa}

   \subtitle{}

   \author{I. Salmaso\,\orcidlink{0000-0003-1450-0869}
          \inst{1}\fnmsep\inst{2}\fnmsep\thanks{\email{ irene.salmaso@phd.unipd.it}}
          \and
          E. Cappellaro \inst{2} \and L. Tartaglia  \inst{2}
          \and S. Benetti  \inst{2} \and M. T. Botticella \inst{3}  \and N. Elias-Rosa  \inst{2}\fnmsep \inst{4} \and A. Pastorello\,\orcidlink{0000-0002-7259-4624}  \inst{2} \and F. Patat \inst{5} \and A. Reguitti\,\orcidlink{0000-0003-4254-2724}  \inst{6}\fnmsep\inst{7}\fnmsep\inst{2} \and L. Tomasella  \inst{2} \and G. Valerin  \inst{1}\fnmsep\inst{2} \and S. Yang \inst{8}
          }

   \institute{Dipartimento di Fisica e Astronomia ``G. Galilei'', Universit\`a degli Studi di Padova, Vicolo dell’Osservatorio 3, 35122 Padova, Italy
         \and
           INAF-Osservatorio Astronomico di Padova, Vicolo dell’Osservatorio 5, 35122 Padova, Italy
           \and INAF-Osservatorio Astronomico di Capodimonte, Salita Moiariello 16, 80131 Napoli, Italy
           \and Institute of Space Sciences (ICE, CSIC), Campus UAB, Carrer de Can Magrans s/n, 08193 Barcelona, Spain
           \and European Organisation for Astronomical Research in the Southern Hemisphere (ESO), Karl-Schwarzschild-Str. 2, 85748 Garching b. München, Germany
           \and Instituto de Astrof\`isica, Departamento de Ciencias F\'isicas -- Universidad Andres Bello, Avda. Rep\'ublica 252, 8320000, Santiago, Chile
           \and Millennium Institute of Astrophysics, Nuncio Monsenor S\'otero Sanz 100, Providencia, 8320000, Santiago, Chile
           \and Department of Astronomy, The Oskar Klein Center, Stockholm University, AlbaNova 10691, Stockholm, Sweden
             }

   \date{Received 23 December 2022; accepted 18 March 2023}

 
  \abstract
   {The link between the fate of the most massive stars and the resulting supernova (SN) explosion is still a matter of debate, in major part because of the ambiguity among light-curve powering mechanisms.
   When stars explode as SNe, the light-curve luminosity is typically sustained by a central engine (radioactive decay, magnetar spin-down, or fallback accretion). However, since massive stars eject considerable amounts of material during their evolution, there may be a significant contribution coming from interactions with the previously ejected circumstellar medium (CSM). 
  Reconstructing the progenitor configuration at the time of explosion requires a detailed analysis of the long-term photometric and spectroscopic evolution of the related transient. 
   }
   {In this paper, we present the results of our follow-up campaign of SN~2020faa. Given the high luminosity and peculiar slow light curve, it is purported to  have a massive progenitor. We present the spectro-photometric dataset and investigate different options to explain the unusual observed properties that support this assumption.}
   {We computed the bolometric luminosity of the supernova and the evolution of its temperature, radius, and expansion velocity. We also fit the observed light curve with a multi-component model to infer information on the progenitor and the explosion mechanism.}
   {Reasonable parameters are inferred for SN~2020faa with a magnetar of energy, $E_p = 1.5^{+0.5}_{-0.2} \times 10^{50}$ erg, and spin-down time, $t_{spin}=15\pm1$~d, a shell mass, $M_{shell} = 2.4^{+0.5}_{-0.4}\,\mathrm{M}_\sun$, and kinetic energy, $E_{kin}(shell) = 0.9^{+0.5}_{-0.3} \times 10^{51}$ erg, and a core with  $M_{core} = 21.5^{+1.4}_{-0.7}\,\mathrm{M}_\sun$ and $E_{kin}(core) = 3.9^{+0.1}_{-0.4} \times 10^{51}$ erg. In addition, we need an extra source to power the luminosity of the second peak. We find that a hidden interaction with either a CSM disc or  several delayed and choked jets is a viable mechanism for supplying the required energy to achieve this effect. }
   {}


   \maketitle
%
\section{Introduction}
\label{sec:intro}

The fates of the most massive stars at $\mathrm{M}>25\,\mathrm{M_\sun}$ remain unclear. Depending on the initial parameters (mass, metallicity, rotation) and configuration (binary separation and mass ratio), they have been purported to produce very different types of supernovae (SNe) and other transients, namely, stripped-envelope SNe~Ic associated to GRBs, SLSNe, multiple outbursts such as SN~2009ip, and (still to be confirmed) pair-instability SNe or even dark collapses \citep{heger2003,langer2022}. Understanding the outcome of massive star evolution is important for studying their contribution to chemical enrichment of the early Universe, since they are the first to explode as SNe and, therefore, to pollute the interstellar medium \citep{schneider2002}. 
At the same time, in the multi-messenger context, there is a strong interest in constraining the nature of the compact remnant (neutron star or black hole) because they contribute to the population of massive black holes observed through gravitational waves \citep{stevenson2022}. 

A key ingredient determining the evolution of massive stars is their mass loss. In fact, during its life, a massive star can shed the outer layers of its envelope through stellar winds or impulsive eruptions. The material ejected in this way piles up around the star and forms the so-called circumstellar medium (CSM). When the star ends its life as SN, the core of the star collapses to a neutron star or to a black hole \citep{woosley2005,janka2012}, with the production of a huge neutrino flux \citep{vartanyan_bh_2023}. The deposition of a small fraction of neutrinos into the inner envelope after core bounce revives the shock wave that powers the envelope ejection with velocity of the order of $10^4\,{\rm km}\,{\rm s}^{-1}$ \citep[][and references therein]{janka_handbook}. The shock break-out (SBO) at the stellar surface causes a short luminosity peak depending on the initial radius of the star \citep{waxman_sbo_handbook}. After that, the luminosity is powered by the radioactive decay chain $^{56}$Ni-$^{56}$Co-$^{56}$Fe and, for SNe with a large H envelope, also by recombination of H, with timescales of $\sim 100$ days. However, if at the time of explosion the CSM material is still confined close to the progenitor star, it can be quickly caught by the fast ejecta and the ejecta-CSM interaction converts part of the kinetic energy of the ejecta into radiation. This mechanism may become the dominant source of luminosity and the SN can thus remain visible for subsequent months or even years \citep{Smith_interaction_handbook_2017}.

Depending on its intensity and geometry, interaction masks what is happening inside the CSM cocoon, making the identification of the progenitor properties very difficult, especially with regard to its mass and the actual explosion mechanism. In fact, in a few cases, SNe that initially appeared as Type IIn were recognised to be Type Ia thermonuclear explosions occurring inside a dense CSM \citep{dilday2012,silverman_ptf11kx_2013}. Nevertheless, we expect that strong interaction occurs often (albeit not exclusively) after the collapse of (very) massive stars ($>30\mathrm{M_{\odot}}$) because these are the most likely to experience strong mass loss or eruptive shell ejections during their late evolution \citep{vink_massloss_2015,smith_LBV_2017}. Hence, studying interacting SNe can connect the observed properties of these transients to their progenitors, shedding light on the link between massive stars and SN explosions.

Also, strongly-interacting SNe are interesting in a multimessenger context, since they are among the candidate sources of high-energy neutrinos up to 1 PeV \citep[][and references therein]{fang2020}, produced in $p-p$ or $p-\gamma$ interactions and subsequent particle decay within the shocked regions. While no direct evidence linking interacting SNe to high-energy neutrinos has been found thus far \citep{fang2020}, a thorough study of these events can help understand whether the physical conditions of the environment may favour cosmic particle acceleration and neutrino production.

Historically, interaction has been the basis of hypothesises seeking to explain the strong radio emission of the Type~II~Linear SN~1979C, but it was then closely associated with Type~IIn SNe, which show narrow Balmer lines in their spectra \citep{schlegel1990}. The persistence of narrow emissions on a SN spectrum is indicative of the presence of a slow-moving CSM ionised by an ejecta-CSM interaction. These transients are generally more luminous than normal Type~II SNe and show a slower luminosity evolution. With regard to SN~2010jl, a well-observed SN~IIn, it reached a maximum bolometric luminosity of $\sim 3 \times 10^{43}\,{\rm erg}\,{\rm s}^{-1}$ and a total radiated energy  $>6.5\times 10^{50}\,{\rm erg}$ \citep{fransson_2010jl_2014}.

 Nonetheless, there are a few cases of SNe with no narrow lines, for which strong interaction was called in to explain the excess of luminosity to supplement the initial ejecta thermal energy and the input from radioactive decays \citep{Smith_interaction_handbook_2017}. In particular, CSM interaction was proposed as powering the high luminosity of Type~II~SLSNe \citep{Kangas2022}. Nevertheless, we stress that the observation of an extra amount of energy in the light curve does not automatically imply that interaction is present: other possible power sources include: energy transfer from the spin-down of a newly-born magnetar \citep{kasen_magnetar_2010} or the late-time fallback of matter onto a black hole \citep{dexter_fallback_2013}. An elevated peak luminosity combined with a long-lasting light curve is often associated with high ejecta mass and, thus, with a massive progenitor. Nevertheless, interaction can contribute to it for a long time and it is almost always expected to happen in SNe from massive stars. A key to identifying SNe with massive progenitors even in a context with interaction is the presence and long-time persistence of spectral lines with broad P-Cygni profiles. In fact, these features are indicative of massive, fast-expanding ejecta above a photosphere.

 In this respect, a compelling case is represented by iPTF14hls, a transient that remained bright for more than 600 days, showing a light curve with at least five peaks; over this entire period, the spectrum showed broad P-Cygni profiles and no apparent decrease in velocity \citep{arcavi_iPTF14hls_2017}. The unique light curve and spectral properties, as well as the huge total radiated energy ($2.20 \times 10^{50}\,{\rm erg}$), could not be adequately explained by any current theory. Possible proposed alternatives include: input from a newly born magnetar \citep{Dessart2018}, disc accretion onto a black hole \citep{chugai_iptf_2018}, long-term outflows from a very massive star \citep{Moriya2020}, and a pulsational pair-instability SN (PPISN) or CSM interaction in an ordinary SN \citep{Woosley2018}. In particular, the last interpretation gained support after the observation of the double-peak H$\alpha$ profile in the very late time spectrum, which suggests a highly asymmetric CSM, possibly arranged in a disc or a torus \citep{andrews_iptf_2018}. 

In the context of SNe with a massive progenitor, our attention was caught by the case of SN~2020faa; based on observations of the first $\sim$ 100 days, it was indicated as a possible clone of iPTF14hls \citep{sheng_2020faa}.
In fact, we previously noted that the early evolution of SN~2020faa was reminiscent  of another peculiar SN, OGLE-2014-SN-073 \citep{terreran_2017}, a bright SN II, with a slow luminosity evolution, broad P-Cygni profiles, and an overall slow spectral evolution. Also, in that case, it was difficult to find a consistent explanation for the observed properties, although this transient was proposed as a promising candidate for a pair-instability SN \citep{Kozyreva2018}.
We decided to continue the photometric and spectroscopic monitoring of SN~2020faa to probe its long-term evolution.

The paper is structured as follows. In Sect.~\ref{sec:observations}, we provide general information on SN~2020faa. In Sect.~\ref{sec:phot}, we present the photometric dataset and we derive the bolometric luminosity in Sect.~\ref{sec:bolom}. We introduce the spectroscopic dataset and relative analysis in Sect.~\ref{sec:spec}. We discuss some relevant properties of the host galaxy in Sect.~\ref{sec:host} and we consider the similarities and differences of SN~2020faa with other SNe in Sect.~\ref{sec:confronto}. We model the bolometric light curve in Sect.~\ref{sec:nagy} and discuss possible interpretations in Sect.~\ref{sec:interpr}.

\section{Observations} \label{sec:observations}

SN~2020faa was discovered on 2020 March 24 (MJD$=58932.604$)\footnote{https://www.wis-tns.org/object/2020faa/} by the Asteroid Terrestrial-impact Last Alert System  \citep[ATLAS]{tonry_atlas_2018,smith_atlas_2020} at coordinates RA=14:47:09.469, DEC= +72:44:11.56. It is located close to a galaxy, WISEA J144709.05+724415.5, also detected  by GALEX as a bright UV source. 
A spectrum of the transient, taken with the Liverpool telescope 12~d after discovery, shows broad H~I and He~I lines on top of a blue continuum and therefore the transient was classified as a Type II SN \citep{Perley2020}. As we discuss in Sect.~\ref{sec:host},  strong narrow emissions of H~I, [O~III], and [S~II] also appear in the spectra at all epochs, originating from a background starburst region.

\cite{sheng_2020faa} (hereafter, Y21) presented the evolution of this transient in the first few months. They focused on the similarity of this event with the extraordinary SN~iPTF14hls, especially given the luminosity evolution in the first six months, with at least two broad peaks, and the spectral appearance, with the persistence in the same period of H I Balmer lines with broad P-Cygni profiles. We frequently refer to their results throughout this paper.

Averaging the position of the narrow lines in our best resolution spectra (cf. Sect.~\ref{sec:spec}), we measured a heliocentric redshift $z=0.03888\pm0.00008$, slightly different from the value reported in Y21 ($z = 0.04106$).
From the measured redshift, after correction to the V3K reference frame ($+20\,{\rm km}\,{\rm s}^{-1}$) and assuming the Planck2018 cosmology ($H_0 = 67.4\,{\rm km}\,{\rm s}^{-1}\,{\rm Mpc}^{-1}$, \citep{planck2018}, we derived a distance modulus $m-M=36.24\pm 0.15$, where the error includes the uncertainty on the galaxy peculiar motion.

We obtained an estimate of the Galactic extinction, $A_V=0.067,$ from the NASA/IPAC Extragalactic Database \citep{schlafly_reddening_2011}. 
At the same time, there is no evidence of extinction occurring in the host galaxy.
In particular, no Na I D narrow absorption is detected in our spectra at the galaxy redshift. 
Hereafter, we correct for extinction assuming only the contribution of the Galactic component.

The field of the SN was monitored by different surveys (ATLAS, ZTF, PS1) during the latter years with no evidence of pre-discovery outburst(s). The latest non-detections before discovery are: i) a relatively shallow upper limit on MJD 58918.59  (ATLAS, filter $orange\,\rm{mag}>18.67$) and ii) a couple of deep limits between three and four days earlier by ZTF, MJD 58915.47 (filter $g\,\rm{mag}>19.37$) and by ATLAS, MJD 58914.64 (filter $orange\,\rm{mag}>19.56$)\footnote{The ATLAS photometry was retrieved from their `forced photometry' server, https://fallingstar-data.com/forcedphot/.}.
In the following, we take the mid epoch between the last nondetection and the discovery date as reference explosion epoch, that is MJD 58926, with an adopted uncertainty of $\pm 7$~d.

We began our photometric and spectroscopic monitoring of SN~2020faa about 200 days after discovery and continued until its luminosity fell below our detection limit,  at about 470 days after discovery.
For the observing campaign, we used a number of telescopes and instruments following the requirement dictated by the declining target luminosity, namely: the robotic Asiago Schmidt telescope + Moravian CCD camera and the Asiago 1.82m telescope + AFOSC~\footnote{https://www.oapd.inaf.it/sede-di-asiago/telescopes-and-instrumentations}, the Liverpool telescope + IO:O~\footnote{https://telescope.livjm.ac.uk/}, the NOT 2.56m + ALFOSC~\footnote{http://www.not.iac.es/}, and the GTC 10.4m + OSIRIS~\footnote{http://www.gtc.iac.es/}.

\section{Photometry} \label{sec:phot}

We used standard reduction techniques to correct the raw images for bias and flat field. Hereafter, we measured the magnitude of the SN with the {\em ecsnoopy} package\footnote{ecsnoopy is a python package for SN photometry using PSF fitting and/or template subtraction developed by E. Cappellaro. A package description can be found at http://sngroup.oapd.inaf.it/ecsnoopy.html.}. 
Template subtraction was necessary for this transient because of the significant contamination from the host galaxy light. To this aim we used as a references the publicly available Pan-STARRS\footnote{https://outerspace.stsci.edu/display/PANSTARRS/} images with the proper filter. With {\em ecsnoopy} we secured the registration of the template image to the same pixel grid of the science image and then used the code {\em hotpants} \citep{Becker2015} for convolution of the two images to the same Point Spread Function (PSF) and photometric scale.
The residual source in the difference image was measured through PSF fitting, which is found to be less sensitive to background noise than plain aperture photometry. If no source is detected above a threshold of 2.5 times, the background noise, a corresponding upper limit is registered. 
Finally, the instrumental magnitudes (or upper limits) were calibrated using photometric zero points measured from local stars with photometry in the Pan-STARRS catalogue\footnote{https://catalogs.mast.stsci.edu/panstarrs/} and using the nominal colour terms for the specific instrument. These photometric measurements are reported in Table~\ref{tab:multiw}.

\begin{table}[htbp]
\caption{Photometry of SN~2020faa measured on our images (AB system).}\label{tab:multiw}
\begin{tabular}{ccrcc}
\hline
    MJD & Band & mag & err & Instrument\\\hline
        59183.159&      i &     20.296  & 0.065&        1.82m+AFOSC   \\ 
        59183.166&      r &     20.445  & 0.067&        1.82m+AFOSC   \\
        59197.940&      r &     20.798  & 0.117&        Schmidt/Asiago \\
        59197.940&      g &     $>21.803$&  (...)   &   Schmidt/Asiago \\
        59198.260&      g &     22.138  & 0.078&        NOT+ALFOSC    \\
        59198.260&      r &     20.800  & 0.066&        NOT+ALFOSC    \\
        59198.260&      i &     20.644  & 0.040&        NOT+ALFOSC    \\
        59232.175&      i &     20.739  & 0.047&        NOT+ALFOSC    \\
        59232.175&      g &     22.164  & 0.080&        NOT+ALFOSC    \\
        59232.175&      r &     21.387  & 0.071&        NOT+ALFOSC    \\
        59260.265&      i &     21.130  & 0.065&        NOT+ALFOSC    \\
        59260.265&      r &     21.528  & 0.082&        NOT+ALFOSC    \\
        59260.265&      g &     $>21.811$&  (...)    &        NOT+ALFOSC\\
        59289.225&      g &     22.197  & 0.094&        NOT+ALFOSC    \\
        59289.225&      r &     21.449  & 0.073&        NOT+ALFOSC    \\
        59289.225&      i &     21.472  & 0.065&        NOT+ALFOSC    \\
        59322.997&      r &     21.598  & 0.059&        NOT+ALFOSC    \\
        59322.997&      i &     21.573  & 0.087&        NOT+ALFOSC    \\
        59322.997&      g &     22.310  & 0.097&        NOT+ALFOSC    \\
        59346.948&      g &     $>22.454$&  (...)   &   GTC+OSIRIS            \\     
\hline
\end{tabular}
\end{table}

A number of optical/UV exposures were obtained with the UVOT of the SWIFT satellite between 100 and 300 d from discovery, which were analysed as follows. We first measured aperture magnitude with centre at the SN position, a radius of 4 arcsec, and the sky background measured in an offset empty region.  We found that in the optical bands (U,B,V) the flux shows an initial decline and then flattens at phases later than 200 d. The flux in the UV filters, instead, remains flat at all epochs. As mentioned above, the transient is projected on a bright H II region, which appears to dominate the integrated flux measurements at late phases. We therefore built template images for each filter by summing the four exposures obtained after MJD 59189 (phase range 264-289 d). After a subtraction of these templates from the images obtained at epochs earlier than 200 d, the transient is detected in all the optical bands but remains below the detection limits in the UV bands. As a further attempt, for the UVW1, UVW2, and UWM2 filters we summed the six exposures obtained before MJD 59054 (phase range 108-129 d). Even in these combined deep images, the transient remained undetected and  we could only obtain more stringent upper limits (cf. Tab.~\ref{tab:swift}). These limits, however, set useful constraints on the spectral energy distribution (SED).

\begin{table}[htbp]
\caption{SWIFT UVOT photometry (AB system)}\label{tab:swift}
\begin{tabular}{ccrcc}
\hline
    MJD & filter & mag & err& \\\hline
59033.924 & U & 20.176 & 0.174 \\
59033.925 & B & 18.384 & 0.050 \\
59033.928 & V & 18.196 & 0.113 \\
59038.698 & U & 19.600 & 0.071 \\
59038.699 & B & 18.410 & 0.046 \\
59038.703 & V & 17.953 & 0.079 \\
 59043.750 & UVM2 & $>21.465$ & (...) &* \\
59043.750 & UVW1 & $>21.045$ & (...) &*\\
59043.750 & UVW2 & $>22.001$ & (...) &*\\
59043.869 & U & 20.124 & 0.146 \\
59043.870 & B & 18.374 & 0.038 \\
59043.875 & V & 17.947 & 0.050 \\
59048.787 & U & 19.878 & 0.113 \\
59048.788 & B & 18.314 & 0.045 \\
59048.792 & V & 17.823 & 0.071 \\
59053.460 & U & 19.844 & 0.098 \\
59053.461 & B & 18.381 & 0.074 \\
59053.465 & V & 17.787 & 0.056 \\
59094.528 & U & $>19.739$ & (...)\\
59094.528 & B & 19.697 & 0.177 \\
59094.533 & V & 18.454 & 0.079 \\
59102.221 & U & $>20.027$ & (...)\\
59102.221 & B & 19.893 & 0.165 \\
59102.223 & V & 18.666 & 0.093 \\
59104.776 & U & $>20.726$ &  (...)  \\   
59104.777 & B & $>20.203$ &  (...)  \\
59104.780 & V & 18.471 & 0.100 \\
59117.447 & U & $>20.130$ & (...)\\
59117.448 & B & $>18.856$ & (...)\\
59117.451 & V & 19.135 & 0.140 \\
59124.382 & U & $>20.994$ & (...)\\
59124.382 & B & $>19.527$ & (...)\\
59124.385 & V & $>18.718$ & (...)\\
59138.685 & U & $>20.726$ & (...)\\
59138.686 & B & $>20.041$ & (...)\\
59138.690 & V & $>18.618$ & (...)\\
\hline
\end{tabular}

* Measured on the sum of six images obtained in the MJD range 59033-59054
\end{table}

\begin{figure}
    \centering
    \includegraphics[width=\columnwidth]{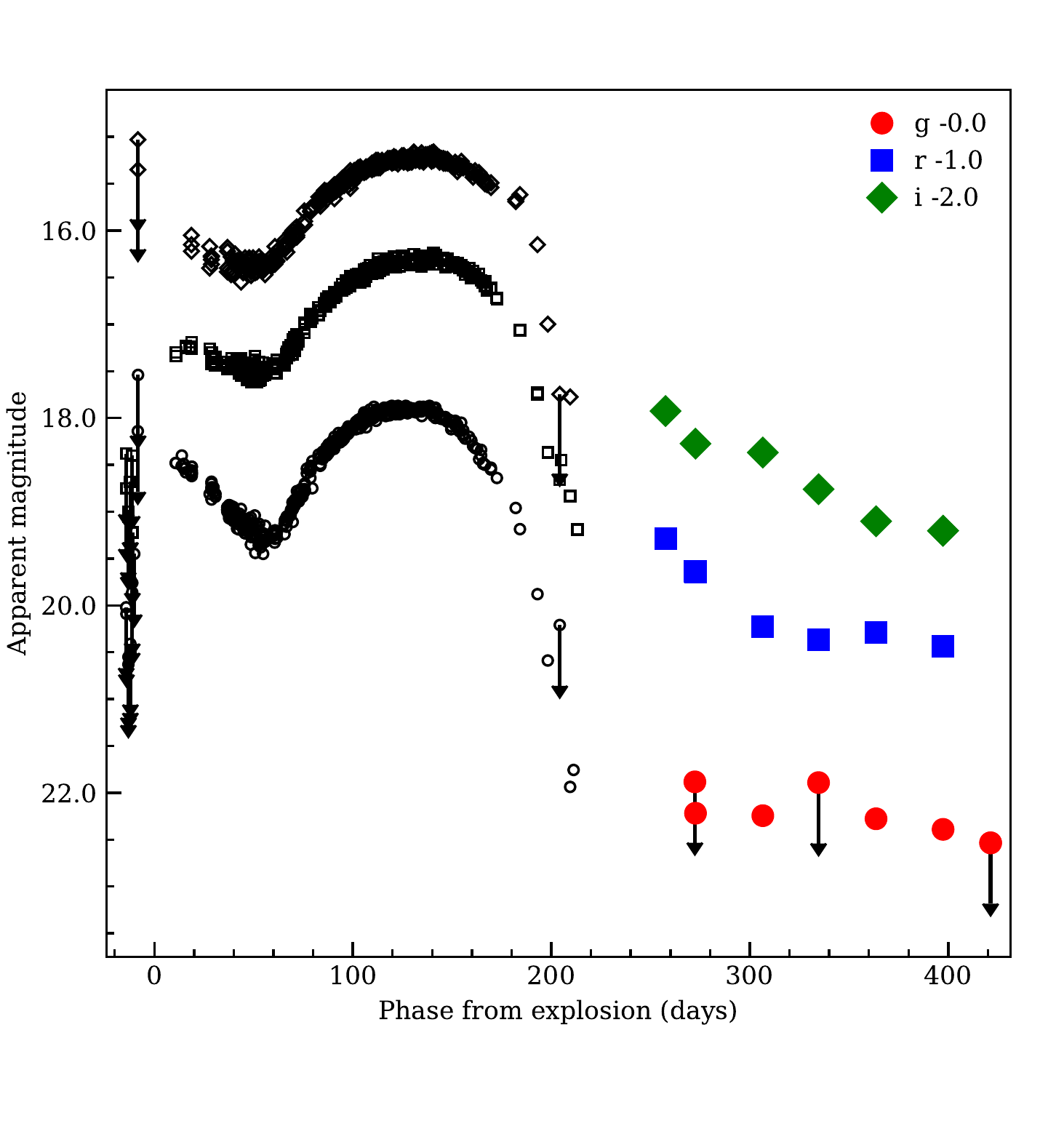}
    \caption{Multi-wavelength light curve of SN~2020faa. Black arrows represent upper limits. Larger, coloured points are the results of our observations (see Tab.~\ref{tab:multiw}), while the others are from Y21.}
    \label{fig:multiwav}
\end{figure}

The full g,r,i light curves of SN~2020faa are shown in Fig.~\ref{fig:multiwav}.
The luminosity evolution appears fairly peculiar with three distinct phases: i) in the first 60 d, a slow, linear decline of 0.5-1.0 mag, slower in r and faster in the g band; ii) between 60 and 200 d, a slow rise to a broad peak. In the r band, the peak occurs at  phase 133 d, with $r=17.27 \pm 0.02$ mag. This is followed by a rapid decline of about 3 mag in $\sim 70$ d; iii) after 250 d and until 400 d, a slow linear declining tail. With the adopted distance modulus and extinction (cf. Sect.\ref{sec:observations}), along with the standard error propagation, we derived a peak absolute magnitude of $M_r=-19.1 \pm 0.2$ mag.

As  pointed out in Y21, up to 150 days, the light curve of SN~2020faa is matched fairly well to that of iPTF14hls. However, at later phases, our new photometry shows a very different evolution, with SN~2020faa monotonically declining, while iPTF14hls remained bright for much longer and showed multiple broad peaks. The result is that while the two SNe have similar peak magnitudes, at 400 days, SN~2020faa is over 4 mag fainter than iPTF14hls.

In some respect, the light curve of SN~2020faa may be taken as a scaled-up version of  SN~1987A \citep{arnett_87A_1989}, with an overall slower evolution and much higher luminosity. 
In particular, if we assume that the late light curve of both SNe is powered by  the $^{56}$Ni$-^{56}$Co$-^{56}$Fe radioactive decay chain, the fact that the luminosity in the tail of SN~2020faa is $\sim 1.5$ mag brighter than for SN~1987A would require a $^{56}$Ni mass that is about three times larger than that of the former SN, namely, over $0.2\,M_{\sun}$. This value is high, but it does not exceed the maximum limit expected for neutrino-driven core collapse of SNe II of $\sim 0.28\,M_{\sun}$ \citep{ni_2017}. However, we go on to show in Sect.~\ref{sec:bolom} that the slope of the tail is slightly slower than the $^{56}$Ni predictions, therefore, we cannot exclude a different power source for the late luminosity. We discuss this further in Sect.~\ref{sec:interpr}. Another SN that was described as a scaled-up version of SN~1987A is OGLE-2014-SN-073 \citep{terreran_2017}. For this SN,  it was also difficult to find a viable explosion mechanism which could explain all the observables. We analyse the similarities between SN~2020faa and OGLE-2014-SN-073 in Sect.~\ref{sec:confronto}.

\section{Bolometric luminosity}
\label{sec:bolom}
The bolometric luminosity ($L_{bol}$) is required for a comparison of the observations with models and, thus, for constraining the physical properties of the source. In principle, deriving the true $L_{bol}$ requires sampling the SED along the full electromagnetic spectrum. Extensive coverage in all photometric bands is rarely available and therefore, as a first-order approximation, it is common practice to integrate the flux over a limited spectral range and then, with some assumptions, estimate a bolometric correction (BC).

In the case of SN~2020faa, extended spectral coverage is available only at few epochs, mostly around the second maximum, while well-sampled light curves are available in just three bands (g,r,i, see Fig. \ref{fig:multiwav}). With this data, we proceeded as follows.

First, we computed a `pseudo-bolometric' light curve integrating the flux included in the g,r,i bands. With this aim, we averaged the measurements obtained with the same filter on the same day (to reduce the light curve scatter), corrected the magnitudes for extinction, and converted them to flux densities using photometric zero points\footnote{http://svo2.cab.inta-csic.es/theory/fps/}. We then integrated the flux in the sampled spectral region using a trapezoidal rule and assuming zero flux below and above the limit defined by the filter equivalent width of the bluer and redder filter, respectively. The measured flux was then converted into luminosity using the adopted distance modulus. The advantage of the gri integration is that it allows for an easier comparison with the same quantity derived for other SNe (see Fig.~\ref{fig:bolom_confronto}).

Nevertheless, there is a significant fraction of flux that is not included in this pseudo-bolometric luminosity and needs to be accounted for modelling purposes.
To devise the BC, a frequently used approach is to fit the SED with a black body function and then integrate it. 
However, it has been shown that for most core-collapse SNe the UV flux is depressed with respect to the black body function due to blanketing from metal lines \citep{dessart_lineblank_2005}, which is stronger when the temperature decreases.  

\begin{figure}
    \centering
    \includegraphics[width=\columnwidth]{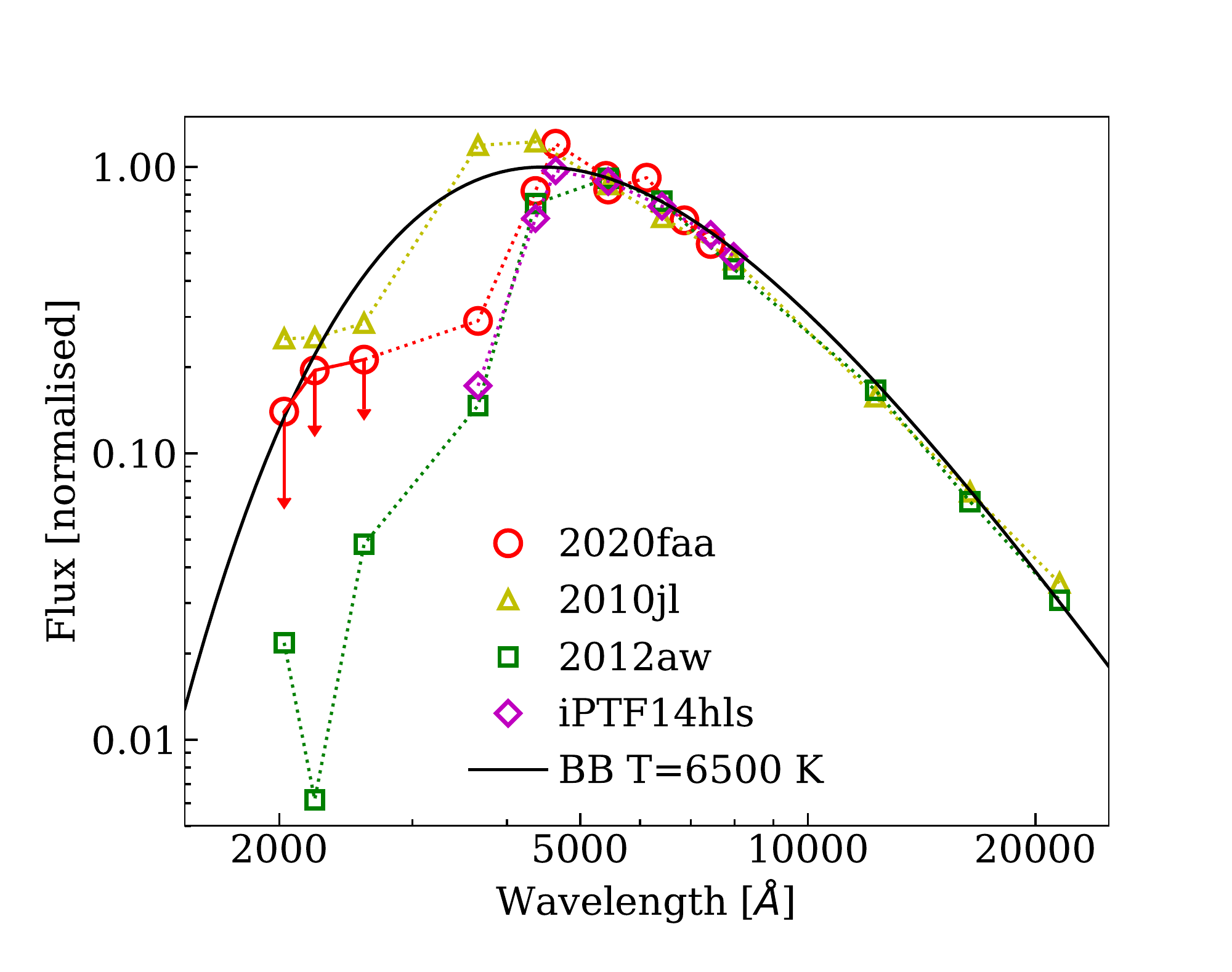}
    \caption{SED of SNe~2020faa (red circles), 2010jl (yellow triangles), 2012aw (green squares), and iPTF14hls (magenta diamonds) compared to a black body with temperature T=6500 K (solid line).}
    \label{fig:sed}
\end{figure}

To check the behaviour of SN~2020faa, in Fig.~\ref{fig:sed} we compared the SED of a few representative SNe at a similar temperature of about $6500\,$K, though at different phases, namely, $\sim130$~d for SN~2020faa, $\sim30$ d for the Type IIn SN~2010jl \citep{fransson_2010jl_2014} and for the standard Type IIP SN~2012aw \citep{dallora2014}, as well as $\sim140$ d for iPTF14hls \citep{arcavi_iPTF14hls_2017}. It appears that the SED of SN~2012aw matches the black body function in the optical and IR, but shows a flux deficiency at shorter wavelengths, while SN~2010jl shows a flux excess in u,g bands. Allowing for a limited spectral coverage, the behaviour of SN~2020faa (and also iPTF14hls) turns out to be similar to that of SN~2012aw, with a blue flux deficiency  already starting in the u band.
Encouraged by the SED comparison, we could exploit the results of \cite{lyman_BC_2014}, who provide a prescription for the BC for regular SNe~II based on measured colours, in particular for g,r bands (their Table~4).

\begin{figure}
    \centering
    \includegraphics[width=\columnwidth]{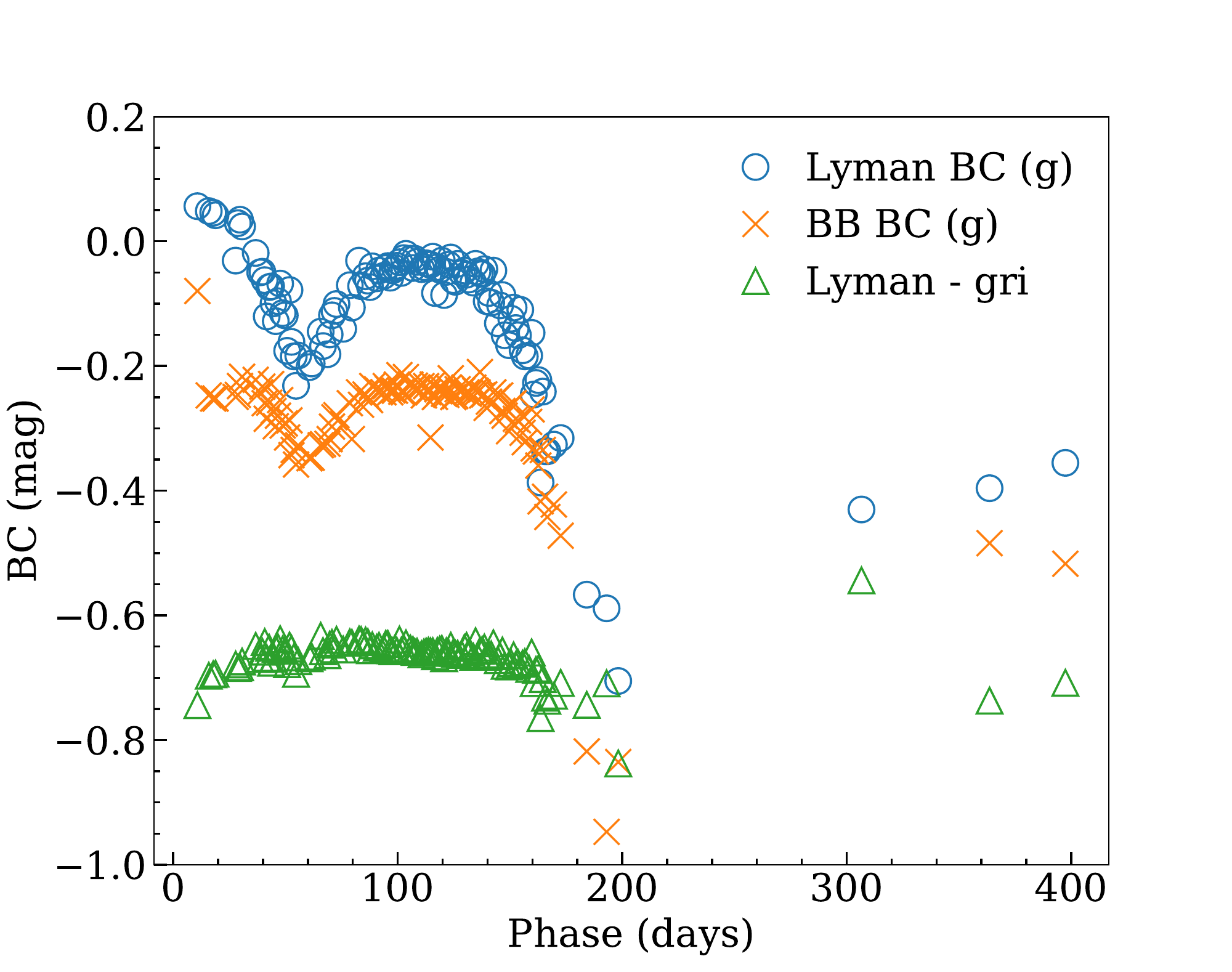}
    \caption{Comparison of different estimates of the bolometric correction relative to the g band. Blue circles are BC derived from \cite{lyman_BC_2014}, while orange crosses are BC after integration of the black body function fitting the available photometry. We also show the difference between Lyman's correction and the gri integration (green triangles).}
    \label{fig:BC}
\end{figure}

A comparison between the different approaches is shown Fig.~\ref{fig:BC}, where we plot the value of BC as a function of time, relative to the g band, for either the black body integration (orange crosses) or Lyman's formulation (blue circles). It can be seen that the two curves have a similar shape but with an offset of about 0.2 mag, which translates to a bolometric luminosity that is about $10$\% fainter when adopting Lyman's prescription with respect to the black body integration.
In the figure, we also show the difference between the gri pseudo-bolometric light curve and the bolometric light curve derived from Lyman's prescription (green triangles). The fact that this difference is fairly constant (0.65 mag) indicates that the gri integration can be taken as a good representation of the bolometric light curve after accounting for the constant offset of 0.65 mag (or 0.26 in $\log L$).

\begin{figure}
    \centering
    \includegraphics[width=\columnwidth]{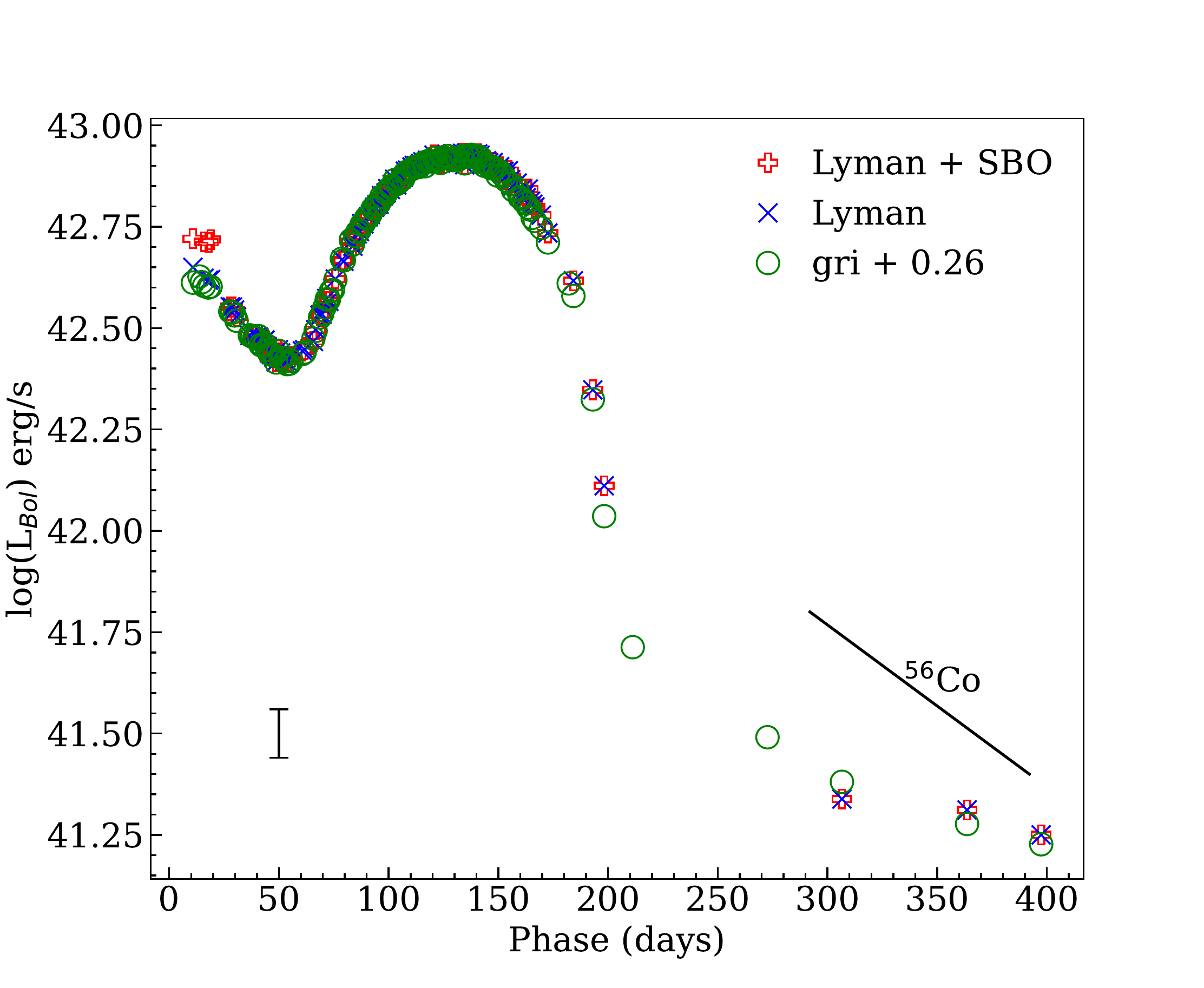}
    \caption{Comparison of the bolometric light curve derived from the gri integration corrected by a constant (+0.26) accounting for the missing flux fraction (open green circles) and Lyman's prescription, assuming different limit for the SBO dominated light curve phase, namely: $g-r<-0.1$ (blue crosses) and $g-r<0.3$ (red plus). The black segment on the lower left of the plot represents the errorbar on absolute luminosity calibration due to the uncertainties on distance modulus and extinction.}
    \label{fig:lbol}
\end{figure}

These differences  are shown in Fig.~\ref{fig:lbol}, where the corrected gri integration is compared with the $L_{bol}$ derived with Lyman's BC with fairly coincident results.
\cite{lyman_BC_2014} noticed that in a fraction of SN~II there is a significant UV contribution from SBO cooling during the early phases that may extend until the colour is bluer than $g-r=0.3$. For these objects, $L_{bol}$ in the early 20 d is $\sim 10$\% higher than the others with the same g-r colour. We have no information to assess whether such correction should be applied to the case of SN~2020faa and thus we chose to neglect it. To visualise this uncertainty, in Fig.~4, the red crosses show the value of $L_{bol}$ assuming the BC with the SBO correction.
 Given these considerations, we elected to choose the corrected gri integration as bolometric light curve to use in the following analysis, in order to best preserve the shape of the light curve for the modelling and represent the data we actually have with minimum assumptions. We note that for the uncertainty of the bolometric luminosity, the contribution of photometric errors is small, of the order $2-3\%$, while we have a larger contribution from the uncertainties on distance and extinction. This latter contribution, however, is independent of the light curve phase and therefore does not affect the shape and slopes of the light curve.

In conclusion, SN~2020faa is a peculiar SN~II whose light curve shows two peaks. The initial, gradual luminosity decline lasting 60~days encompasses a total radiated energy of $\sim 1.25 \times 10^{49}$ erg. This decline is followed by a broad, bright peak with a maximum at $L_{peak}= 8.38 \pm 0.06 \times 10^{42}\,\rm{erg\,s^{-1}}$. Finally, the long-lasting tail with a slope $\sim 0.8 \pm 0.2$ mag/100~d is somewhat slower, though formally consistent, with the expectation for a Ni-powered light curve ($\sim 1$ mag/100~d, \citealt{woosley1989}).

An estimate of the photospheric temperature is derived from the black body SED fitting. Then, by exploiting $L_{bol}$, we derived an estimate of the photospheric radius (Fig.~\ref{fig:TeR}). We discuss this in the context of a comparison to other SNe in Sect.~\ref{sec:confronto}.

\section{Spectroscopy}
\label{sec:spec}

The spectroscopic observations were reduced using standard prescriptions with the package {\em foscgui}\footnote{foscgui is a python/pyraf-based graphic user interface aimed at extracting SN spectroscopy and photometry obtained with FOSC-like instruments. It was developed by E. Cappellaro. A package description can be found at http://sngroup.oapd.inaf.it/foscgui.html}. In short, we corrected for bias and flat field, removed cosmics, calibrated in wavelength the 2D frame, and extracted the 1D spectrum. Then, we calibrated in flux and corrected for second-order contamination (if required) and telluric absorption (in the case of the latter, using the spectrum of a standard star). 

Also, because of its importance, we attempted a new, careful reduction of the classification spectrum obtained on MJD 58945.5 (phase 20 d) with the SPRAT spectrograph at the Liverpool Telescope (D.A. Perley, p.c.). 

\begin{table*}
\caption{Spectra of SN~2020faa in our dataset. Phases are calculate from the assumed explosion epoch (see Sect.~\ref{sec:observations}).}\label{tab:lista_spettri}
\begin{tabular}{cccccccc}
\hline
        Date & MJD      & Phase (d)     &       Telescope & Grism & Slit ('') & Resolution (\AA) & Exposure (s)\\\hline
2020-04-05      & 58945.5 &     +20     &       LT+SPRAT & Wasatch VPH & 1.8 & 18 & 750\\
2020-07-03 &    59033.5 &       +108 &  NOT+ALFOSC & Gr 4 & 1.0 & 17 & 600\\
2020-07-24 &    59055.5 &       +130 &  NOT+ALFOSC & Gr 4 & 1.3 & 18 & 500\\
2020-08-15 &    59077.4 &       +152 &  NOT+ALFOSC & Gr 4 & 1.0 & 14 & 900\\
2020-08-30 &    59092.4 &       +167 &  NOT+ALFOSC & Gr 4 & 1.0 & 14 & 2400\\
2020-09-17 &    59110.3 &       +185    &       Copernico+AFOSC & VPH6+7 & 1.69 & 16 & 2700\\
2020-09-19 &    59112.4 &       +187    &       NOT+ALFOSC & Gr 4 & 1.0 & 15 & 600\\
2020-10-15 &    59138.3 &       +213 &  GTC+OSIRIS & R1000B+R & 1.0 & 7 & 1800\\
2021-01-02 &    59216.8 &       +291 &  GTC+OSIRIS & R1000B & 1.0 & 7 & 3000\\
2021-05-12 &    59347.5 &       +422 &  GTC+OSIRIS & R1000B & 1.0 & 7 & 3600\\
2021-06-30 &    59396.5 &       +471 & GTC+OSIRIS & R1000B & 1.0 & 7 & 3600\\
\hline
\end{tabular}
\end{table*}

\begin{figure}[htbp]
\centering
\includegraphics[width=\columnwidth]{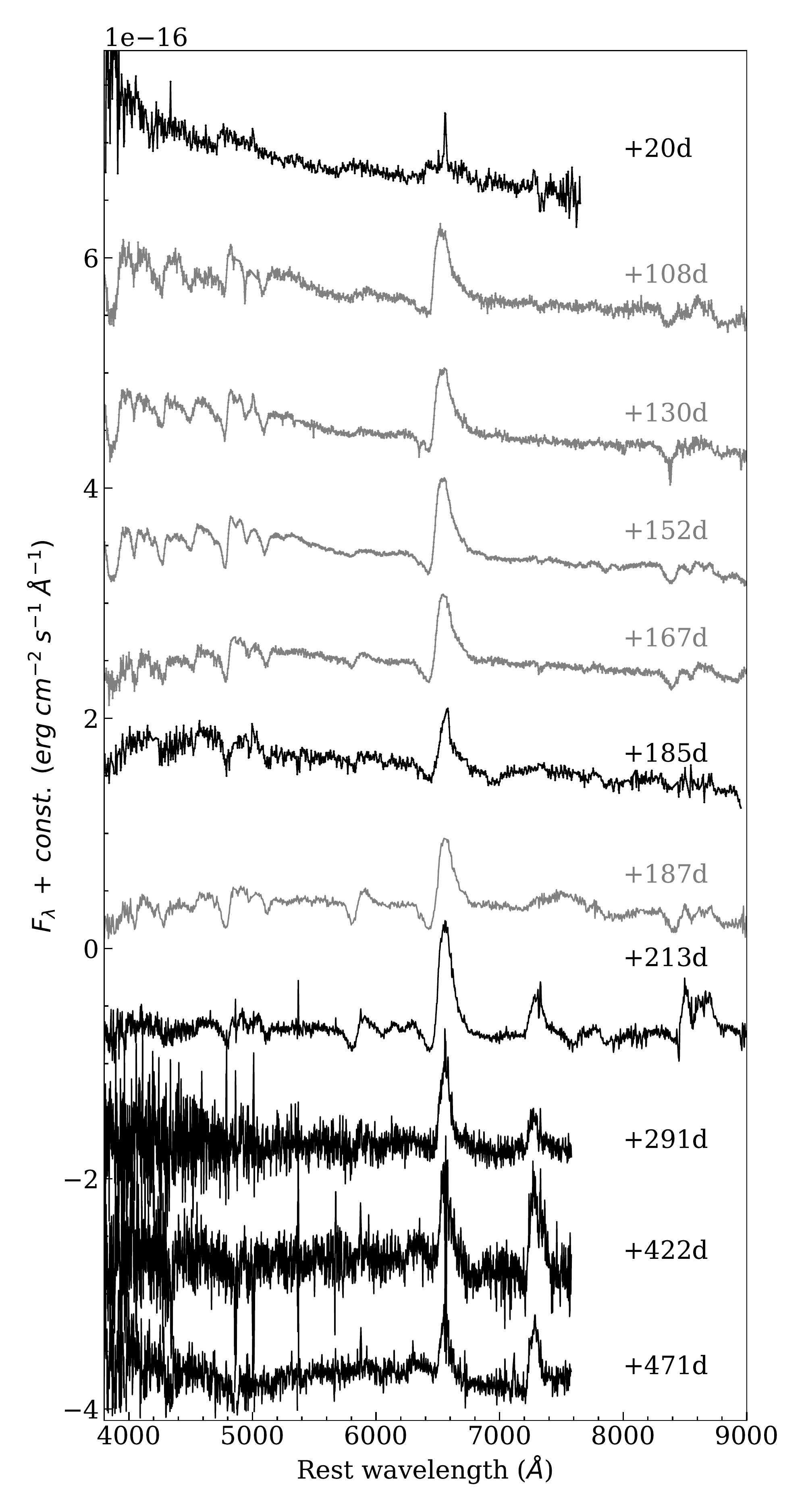}
\caption{Spectral evolution of SN~2020faa. Labels indicate the days after the explosion. Spectra in gray are from Y21, while the spectra in black were reduced by us. All spectra are reddening- and redshift-corrected and matched in intensity with respect to the H$\alpha$ emission and arbitrarily shifted for better visualisation. All spectra are galaxy template-subtracted (see Fig.~\ref{fig:sub} and text).}
\label{fig:spettri}
\end{figure}
    
Our reduced spectra, integrated with the most representative spectra presented by Y21, are listed in Table~\ref{tab:lista_spettri} and shown in Fig.~\ref{fig:spettri}.
The spectra are plotted after correction for the redshift ($z=0.03888$) and for the Milky Way reddening towards the host galaxy ($A_V=0.067)$ (cf. Sec.~\ref{sec:observations}). 

They are dominated by a broad H$\alpha$ with a P-Cygni profile that persists for at least 200~days and appears only as a broad emission after that. 
The other persistent features are narrow H~I, [O~III] $\lambda\lambda 4859,5007$, and [S II] $\lambda\lambda 6716,6731$ emissions, the latter best seen in the latest spectra. As we argue in Sect.~\ref{sec:host}, these emissions originate from a background H~II region. Instead, we find no evidence of evolving narrow line emission possibly associated with circumstellar matter surrounding the SN, with the possible exception of the spectrum at phase +20, which, however, does not have optimal S/N and resolution.

\begin{figure}
    \centering
    \includegraphics[width=\columnwidth]{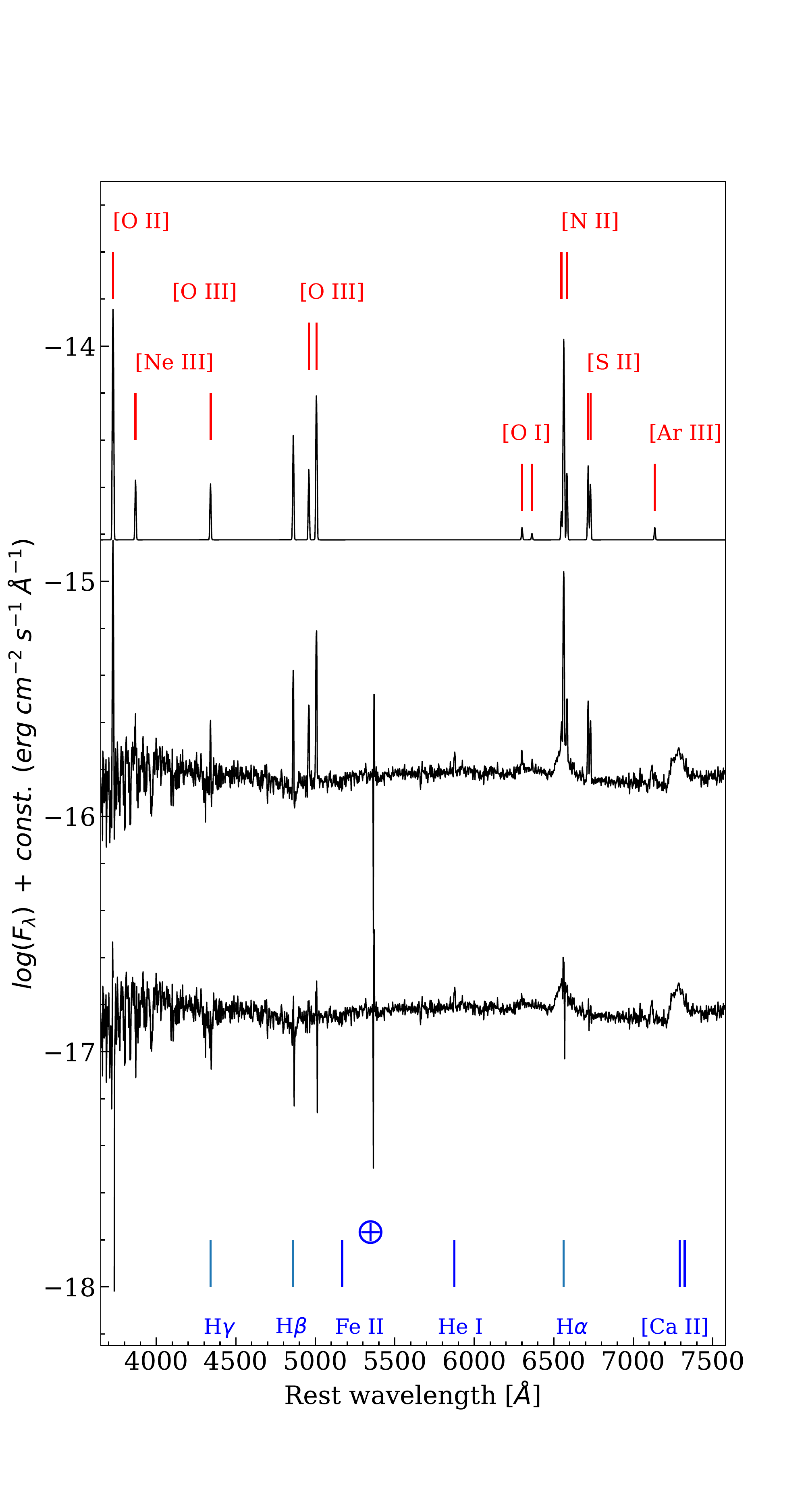}
    \caption{Template subtraction on the spectrum at phase +471 d. The upper spectrum is the template for the line subtraction generated from the middle one (see text), while the lower spectrum is the result of the subtraction. We also indicate the main emission lines. The blue cross indicates a telluric line ([O~I]$\lambda5577$), which was poorly subtracted and is still present in the final result.}
    \label{fig:sub}
\end{figure}

The spectrum of the background source is dominant on the latest spectrum obtained at 471 d, although broad SN features of H~I and [Ca~II] are still detected. In order to highlight the SN emission, we measured the intensity and FWHM of the narrow emission lines in this spectrum and built a synthetic template with the Balmer series, [O~II]$\lambda3727$, [Ne~II]$\lambda3868$, [O~III]$\lambda\lambda\lambda 4342,4959,5007$, [O~I]$\lambda\lambda6300,6364$, [N~II]$\lambda\lambda 6548,6583$, [S~II]$\lambda\lambda 6716,6731$, and [Ar~III]$\lambda7136$. We also added a continuum contribution estimated from a fourth-degree polynomial fit and subtracted the resulting template from the spectrum. The template, the original spectrum, and the resulting subtracted spectrum, along with the line identification, are shown in Fig.~\ref{fig:sub}. The subtracted spectrum of SN~2020faa shows broad emission from the Balmer series and [Ca~II]$\lambda\lambda 7293,7326$, as well as a poorly subtracted telluric line due to [O~I]$\lambda 5577$. We also mark the position of Fe~II $\lambda 5169$ and He~I $\lambda 5875$, even though they are only marginally seen and are more dominant in earlier spectra.
Given the good result of the subtraction, we elected to apply this procedure also to the earlier spectra. The spectra in Fig.~\ref{fig:spettri} are all template-subtracted.

\begin{figure}[htbp]
    \centering
\includegraphics[width=\columnwidth]{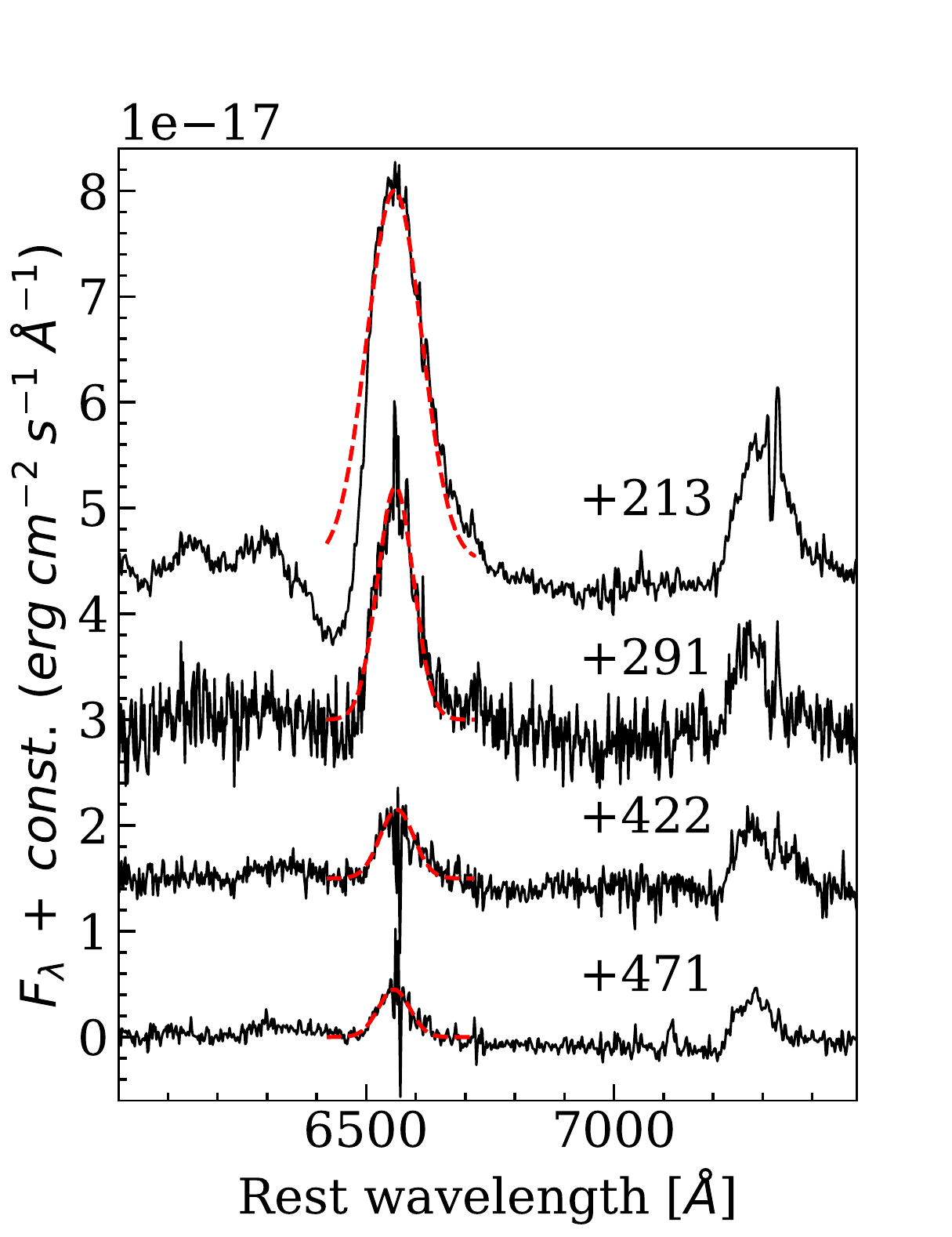}
\caption{Zoom-in on the H$\alpha$ region of the last four spectra. The dashed red line shows how, while in the last three spectra the line profile is approximately Gaussian, the earlier one still shows an asymmetric structure due to the P-Cygni absorption. All spectra are reddening- and redshift-corrected and matched in intensity with respect to the H$\alpha$ emission and arbitrarily shifted for better visualisation.}
    \label{fig:halpha}
\end{figure}

The overall spectral evolution is similar to that of typical type II SNe but for the very slow timescale, with broad P-Cygni progressively shrinking as the ejecta opacity decreases and deeper, slower layers are exposed. While earlier spectra are blue and dominated by Balmer lines, the later ones have a redder continuum and more evident metal lines. Also, Fe~II emissions are present at all phases.

The first spectrum was taken 20~d after the explosion and is the only one in our sample taken during the phase of initial decline. It is unfortunately noisy, but broad Balmer lines are clearly visible. There is also a narrow line on top of the H$\alpha$ which, as mentioned before, could be a contribution of the CSM surrounding the SN, rather than contamination from the host galaxy. Allowing for the low S/N, the broad H$\alpha$ seems to have a boxy profile, which could again be indicative of circumstellar interaction. At around 5875 \AA, we can also see the bump of broad He~I. The continuum is very blue and yields a black-body temperature of $T_{BB}=9000 \pm 1000$ K, fully consistent with the value derived from the fit of the broadband photometry. Spectra taken at phase 108-167~d are all very similar. Their main feature is the H$\alpha$ P-Cygni profile, which maintains a constant velocity of about $6500\,\rm{km\,s^{-1}}$. The black-body temperature decreases as well to $T_{BB}=7200 \pm 500$ K. The spectra at phases 185-187 d show a small bump in correspondence of the blended feature of [Ca~II]$\lambda \lambda 7293,7326$ and O~I $\lambda 7774$, which becomes evident in the spectrum at 213 d. The same happens for the blend of O~I $\lambda8446$ and Ca~II $\lambda\lambda\lambda8498,8542,8662$. 
In these late spectra, the continuum is very faint if any. Moreover, we note that the broad emission lines due to the SN ejecta are progressively shrinking in the last four spectra and the P-Cygni absorption disappears as the radius of the photosphere decreases. In Fig.~\ref{fig:halpha}, we plot a zoom-in on the region of H$\alpha$ for the last four spectra in our sample, along with a Gaussian fit on the emission line. The H$\alpha$ profile at 213 d is still clearly asymmetric due to the P-Cygni absorption cutting off the bluer flux. A small absorption could also be present at 291~d, while the Gaussian profile matches well the H$\alpha$ profiles at phase 422 and 471~d.

\begin{figure}[htbp]
        \centering
        \includegraphics[width=\columnwidth]{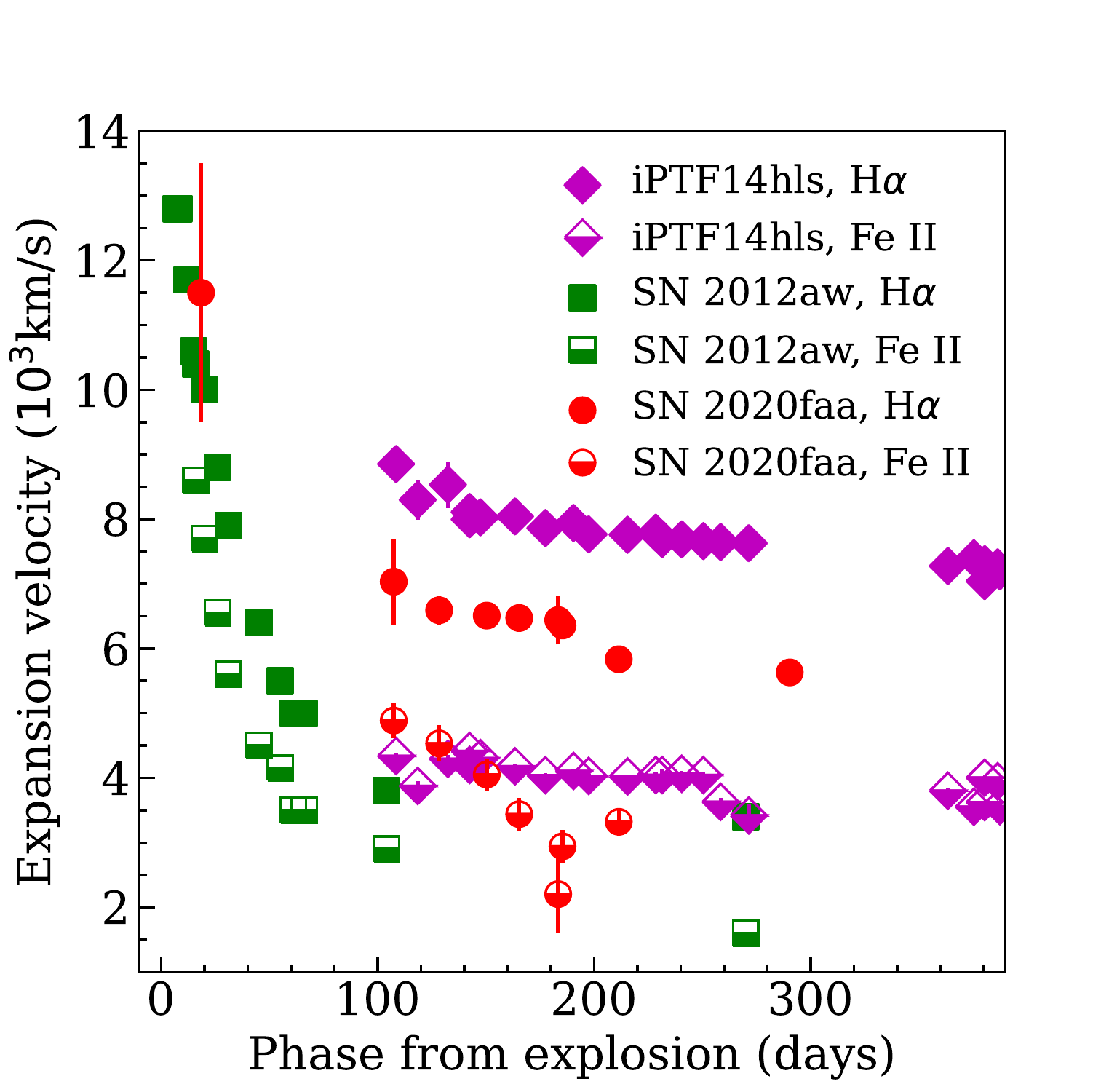}
        \caption{Expansion velocity of SN~2020faa from the H$\alpha$ and Fe~II P-Cygni absorption, compared with iPTF14hls (data from \citealt{arcavi_iPTF14hls_2017}) and SN~2012aw (data from \citealt{bose_2012aw_2013}).}
        \label{fig:vexp}
    \end{figure}

In Table~\ref{tab:vexp}, we report the evolution of the expansion velocity measured from the minimum of the H$\alpha$ and Fe~II P-Cygni absorption. The expansion velocity from H$\alpha$ decreased steadily, from $11500\,\rm{km\,s^{-1}}$ 20~days after the explosion to $5600\,\rm{km\,s^{-1}}$  at 291~d. The velocity from Fe~II follows the same trend, although we could not securely detect the lines in the earliest and latest spectra, relying instead on overall lower values (from 4900 at 108~d to 3300 at 213~d). This is a fairly normal occurrence among SNe II, considering the different optical depths, and it is well illustrated in Fig.~\ref{fig:vexp}, where we plot the evolution of the expansion velocity from both lines in comparison with the aforementioned iPTF14hls and SN~2012aw. While the H$\alpha$ expansion velocity initially resembles that of SN~2012aw, the decrease rate is significantly slower for SN~2020faa after 100~d and is similar to the trend set by iPTF14hls, although the latter does show, at all phases, a higher velocity. On the other hand, in SN~2020faa the velocity from Fe~II decreases slowly but monotonically, while it remains constant in iPTF14hls. The latter is attributed to the line-forming region being detached from the photosphere \citep{arcavi_iPTF14hls_2017}.

\begin{table}
\caption{Expansion velocity of SN~2020faa measured on the P-Cygni absorption of H$\alpha$ and Fe~II $\lambda 5169$.}
\label{tab:vexp}
\begin{tabular}{ccc}
\hline
Phase (d) & $v_{exp}^{H\alpha}\,(km\,s^{-1})$ & $v_{exp}^{Fe\,II}\,(km\,s^{-1})$\\\hline
+20 & $11500 \pm 2000$ & (...) \\
+108 & $7000 \pm 700$ & $4900 \pm       300$\\
+130 & $6600 \pm 200$ & $4500 \pm       300$\\
+152 & $6500 \pm 200$ & $4100 \pm       200$\\
+167 & $6500 \pm 200$ & $3400 \pm       300$\\
+185 & $6400 \pm 300$ & $2200 \pm       600$\\
+187 & $6400 \pm 200$ & $2900 \pm       300$\\
+213 & $5800 \pm 200$ & $3300 \pm       200$\\
+291 & $5600 \pm 200$ & (...) \\
\hline
\end{tabular}
\end{table}

\section{Host galaxy}
\label{sec:host}

\begin{figure}
    \centering
    \includegraphics[width=0.49\columnwidth]{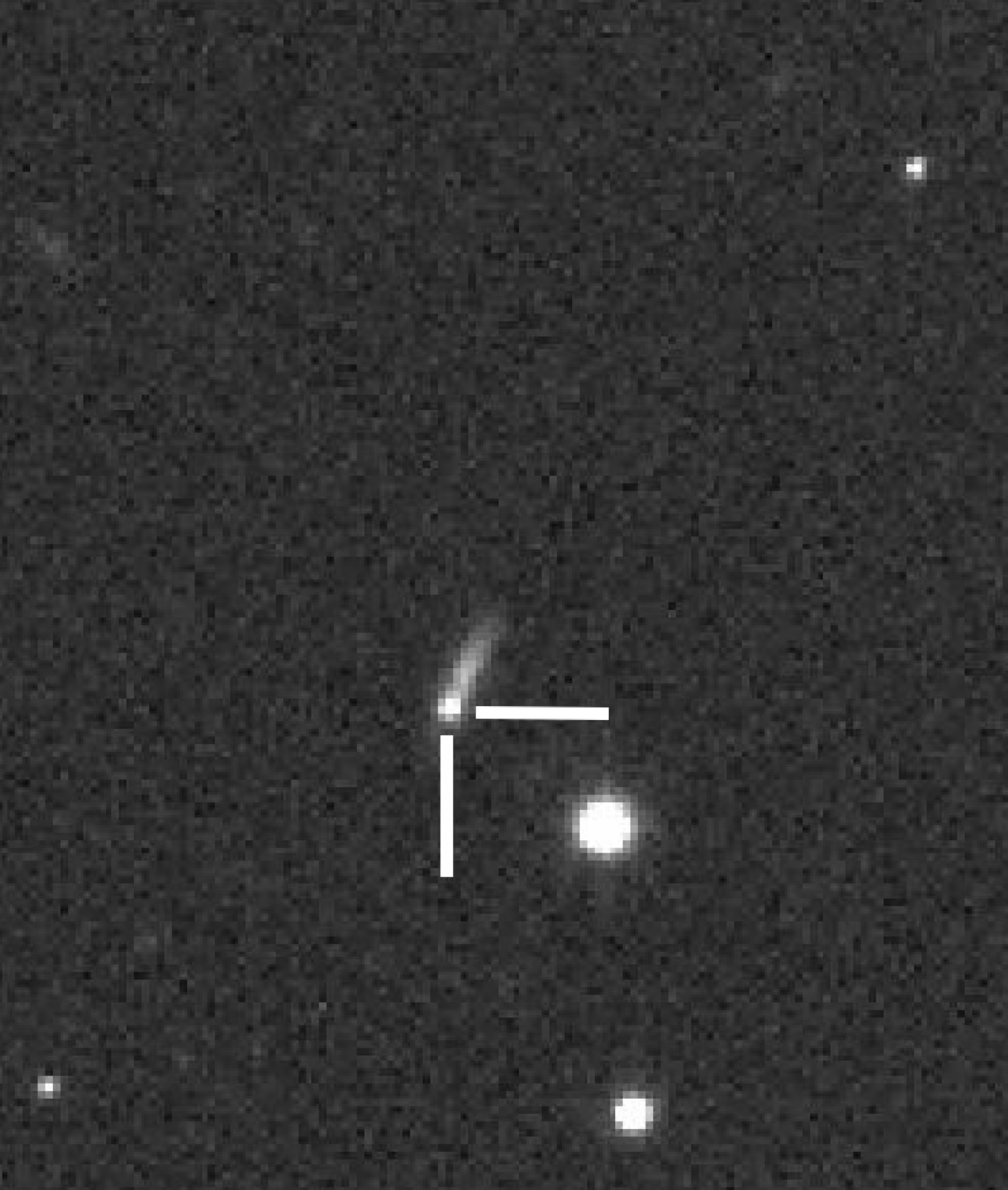}\includegraphics[width=0.49\columnwidth]{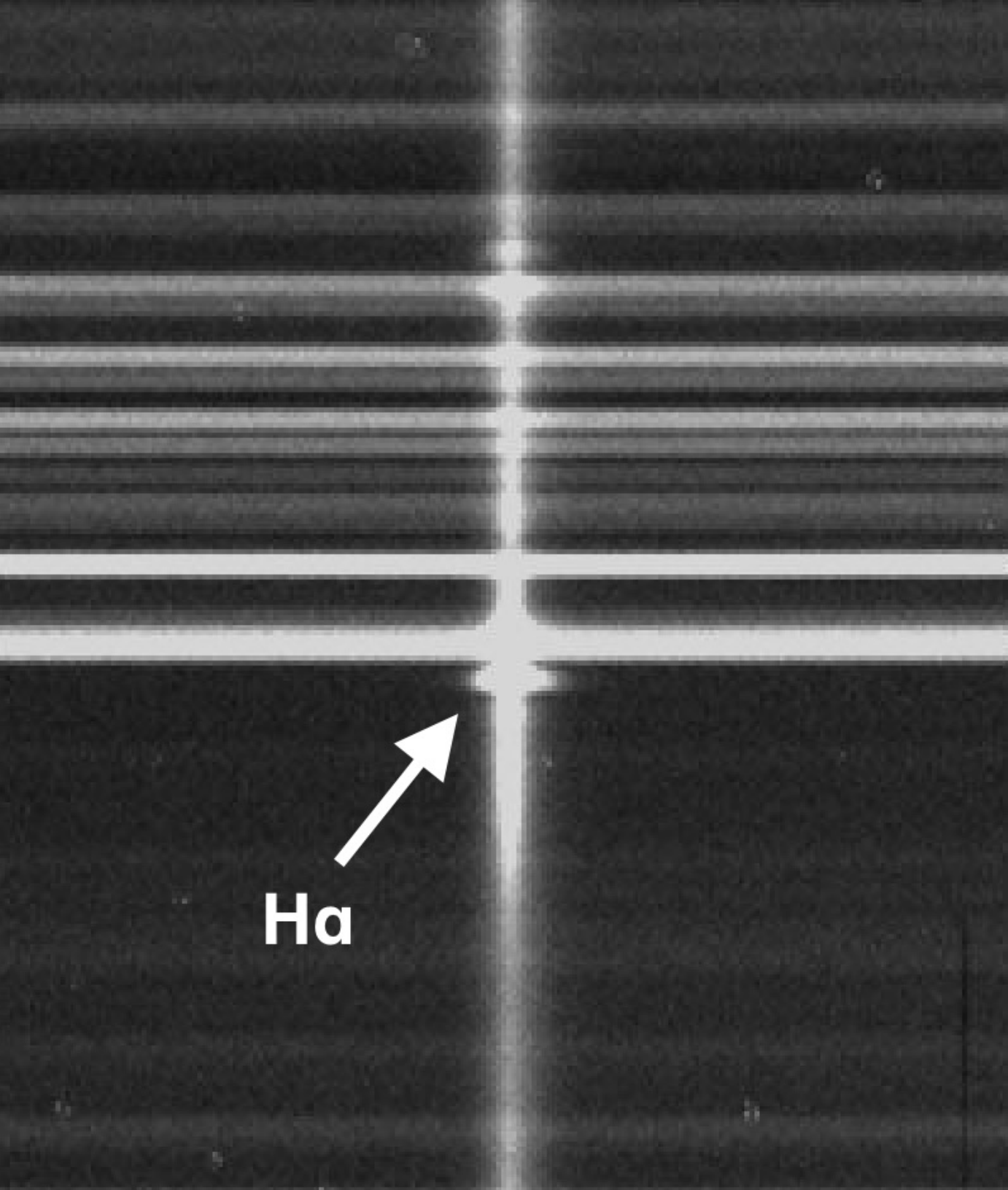}
    \caption{SN~2020faa in imaging and spectrum. \textit{Left:} Finding chart for SN~2020faa. \textit{Right:} Zoom on the H$\alpha$ region of 2D spectrum of SN~2020faa. We notice the extended emission due to the underlying H~II region.}
    \label{fig:fc}
\end{figure}

In Sect.~\ref{sec:spec}, we mention the presence of narrow lines in the spectra that are not due to the SN itself but to the underlying H~II region. The resolution in the photometric images is not enough to resolve the region but its emissions are clearly present in the spectra, as can be seen from Fig.~\ref{fig:fc}.

\begin{figure}
    \centering
    \includegraphics[width=1.2\columnwidth]{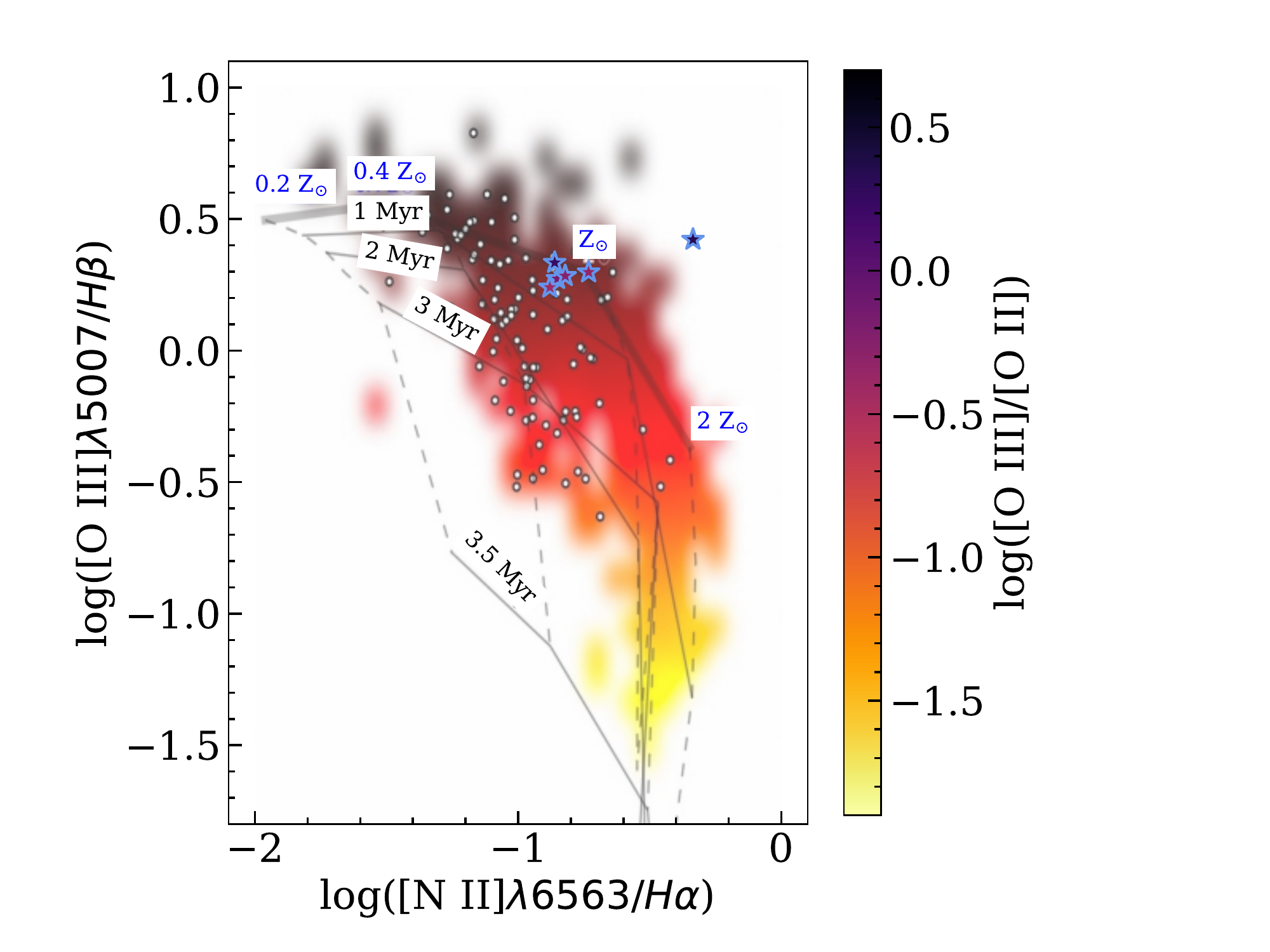}
    \caption{BPT diagram of the underlying H~II region. We plot our measurements (purple stars) over Fig.~5 of \cite{bresolin_HII_2012}. The positioning of the line ratios in the diagram is, for all spectra, consistent with a fairly young, metal-poor region. Solid lines display the models from \cite{dopita2006} for ages $0.1-3.5$ Myr and metallicities $0.2-2\,\rm{Z}_{\sun}$, while dashed lines indicate constant metallicity.}
    \label{fig:hii}
\end{figure}

To characterise the SN environment, we measured the relative intensity of narrow lines ([N~II]$\lambda6583$ and H$\alpha$, [O~III]$\lambda5007,$ and H$\beta$) and overplotted their ratio on the Baldwin, Phillips \& Terlevich (BPT) diagram by \cite{bresolin_HII_2012} in the latest six spectra (see Fig.~\ref{fig:hii}). Here, the solid tracks represent the model grid by \cite{dopita2006} for different ages and metallicity. The measured line ratios are mostly consistent with a H~II region with an age of $\sim1$ Myr and a solar/sub-solar metallicity. There is no significant evolution with time and all data points tend to cluster, except for one of them, which corresponds to the spectrum taken at phase +187~d. This small difference is not surprising and can be due to the effect of a different slit orientation or exact position for background extraction. The fact that, aside from this variation, the measurements are consistent with each other, has pointed us to the conclusion that the narrow lines should be attributed to the H~II region and not to the interaction of the SN ejecta with the CSM.

A SN exploding in an environment at solar metallicity and age 1 Myr gives us information on the progenitor. 1 Myr is a very short time, considering that the expected lifetime for a $40\,M_{\odot}$ is $\sim$5 Myr \citep{chieffi_limongi_2013}, therefore, the progenitor of SN~2020faa is likely a very massive star. This is consistent with the high mass for SN~2020faa that we derive from the light curve modelling in the next section.

\section{Peculiarity of SN~2020faa}
\label{sec:confronto}
\begin{figure}
        \centering
        \includegraphics[width=\columnwidth]{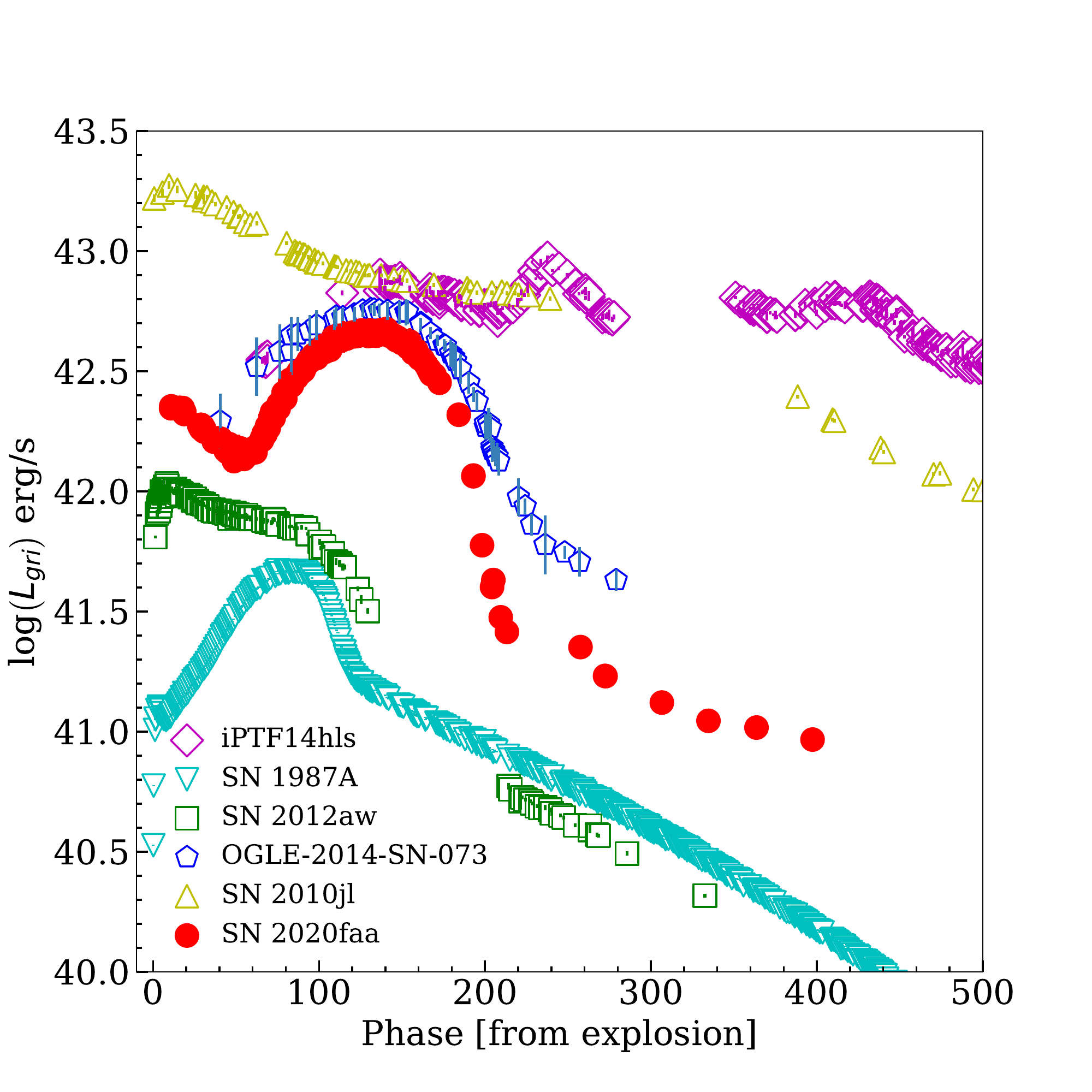}
        \caption{Comparison of SN~2020faa pseudo-bolometric light curve (optical bands only were used in the computation) with other known SNe. All light curves are matched at the moment of explosion but for OGLE-2014-SN-073. Given the uncertainty in the explosion date, we elected to match its peak to the one of SN~2020faa.}
        \label{fig:bolom_confronto}
    \end{figure}

In Fig.~\ref{fig:bolom_confronto}, we compare the pseudo-bolometric gri light curve of SN~2020faa with a number of representative SNe. In order to make them comparable, we calculated the pseudo-bolometric of all SNe from optical bands only with the procedure presented in Sect.~\ref{sec:bolom} and using the same cosmology reported in Sect.~\ref{sec:observations}.

SN~2010jl is one of the most luminous SNe~IIn and it is the brightest SN in this comparison. The peak is significantly brighter than that of SN~2020faa.
On the other hand, SN~2012aw is substantially dimmer than SN~2020faa but, such as SN~2010jl, has a narrow peak close to the explosion date. Given the uncertainty in the explosion epoch of SN~2020faa, it is possible that such a narrow peak has been lost. The following plateau phase of SN~2012aw, instead, is only slightly dimmer than the initial decline of SN~2020faa and has a similar length. This could indicate that it is powered by the SBO cooling of the SN ejecta, which, in the case of SN~2020faa, points toward a larger radius than for SN~2012aw, thus explaining its higher luminosity.
SN~1987A seems a scaled-down version of SN~2020faa, although the duration of the initial decline is much shorter than for SN~2020faa. The decline after the peak is also similar for both SNe, with SN~2020faa declining at a slightly slower rate.
The light curve of iPTF14hls matches really well SN~2020faa up to the peak, but the following evolution is quite different, with the former SN showing multiple peaks stemming from interaction with several shells of CSM \citep{andrews_iptf_2018}.
OGLE-2014-SN-073 appears to be the most similar object in our sample. The peak luminosity is similar, as well as the overall shape of the light curve. In particular, the peak broadness is almost the same and considering the large uncertainty in the explosion date of OGLE-2014-SN-073 \citep{terreran_2017}, they could also have a similar evolutionary timescale. Also, while the tail of OGLE-2014-SN-073 is not well-sampled because the SN disappeared behind the Sun, the last data point at almost 500~d indicates a linear decline that is very similar to that of  SN~2020faa.

\begin{figure}
    \centering
    \includegraphics[width=\columnwidth]{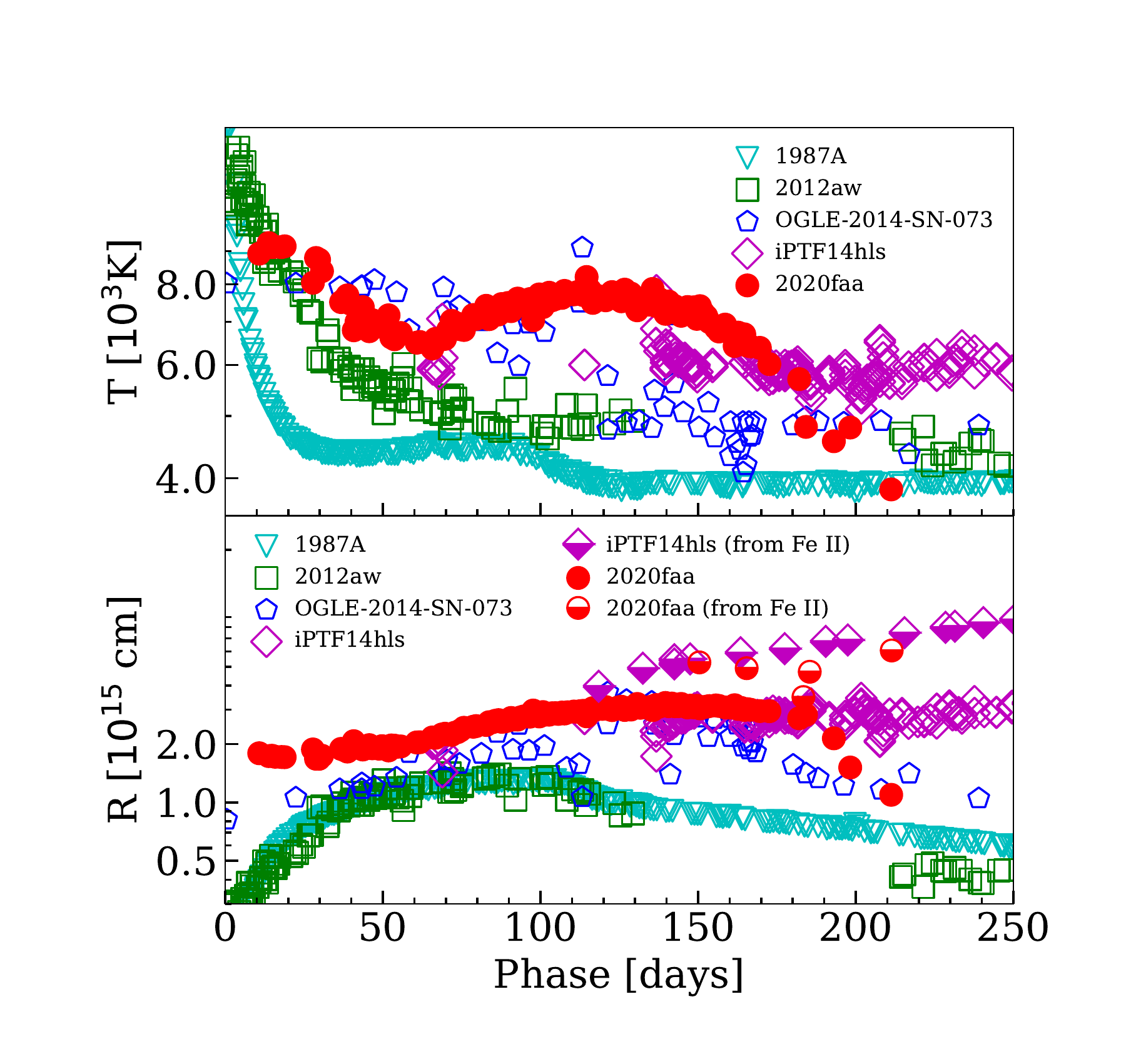}
    \caption{Temperature and radius evolution for SN~2020faa compared with other significant SNe. \textit{Upper panel:} Temperature evolution from black body SED fitting of SN~2020faa compared with SNe~1987A, 2012aw, OGLE-2014-SN-073, and iPTF14hls. \textit{Lower panel:} Photospheric radius of SN~2020faa from black body temperature and luminosity, compared with the same SNe, and kinematic radius calculated from expansion velocity of Fe II $\lambda 5169$ for SNe~2020faa and iPTF14hls.}
    \label{fig:TeR}
\end{figure}

As mentioned in Sect.~\ref{sec:bolom}, we derived the photospheric temperature  as a function of time from the SED fit. In Fig.~\ref{fig:TeR}, we compare the values for SN~2020faa with those of SNe~1987A, 2012aw, OGLE-2014-SN-073, and iPTF14hls. In the early phases, all SNe show a steady decrease in their temperature, as expected by an expanding photosphere, although this is faster for SNe~1987A and also 2012aw to a lesser extent.
The temperature of SN~2020faa, instead, starts with a slow decrease in the first $\sim$60~d and then increases following the trend of the bolometric light curve during the peak. SN~1987A is the only one in this comparison showing a slight increase in the temperature, but not as much as SN~2020faa. This behaviour would point to an extra source of energy in the phase 60-200~d for SN~2020faa.

We also derived the radius from the temperature and luminosity, after the correction for the dilution factor \citep{vogl2019} that accounts for the deviation of the SED from a plain black body function. In this case, we assume that the correction valid for SNe II is applicable to SN~2020faa. The radius of SN~2020faa shows a slow but steady increase until $\sim$200 d, similar to that of iPTF14hls and OGLE-2014-SN-073, since all three SNe start from very large radii. On the other hand, SNe~1987A and 2012aw have a faster increase from a considerably smaller radius until $\sim$50 d. In fact, even with the fast increase, they never reach the size of SN~2020faa radius.
We also calculated the kinematic radius of SN~2020faa as $R_k=v\cdot t$, where \textit{v} is measured from the Fe II $\lambda5169$ P-Cygni absorption (see Sec.~\ref{sec:spec}) and compare it with that of iPTF14hls (data from \citealt{arcavi_iPTF14hls_2017}). For both SNe, the radius thus calculated does not match the one from bolometric luminosity, even with the correction for the dilution factor. \cite{arcavi_iPTF14hls_2017} attributed the observed difference in iPTF14hls to a line-forming region that is detached from the photosphere; however, the question of whether this is a viable explanation  for SN~2020faa as well is unclear.
    
\begin{figure}
        \centering
        \includegraphics[width=\columnwidth]{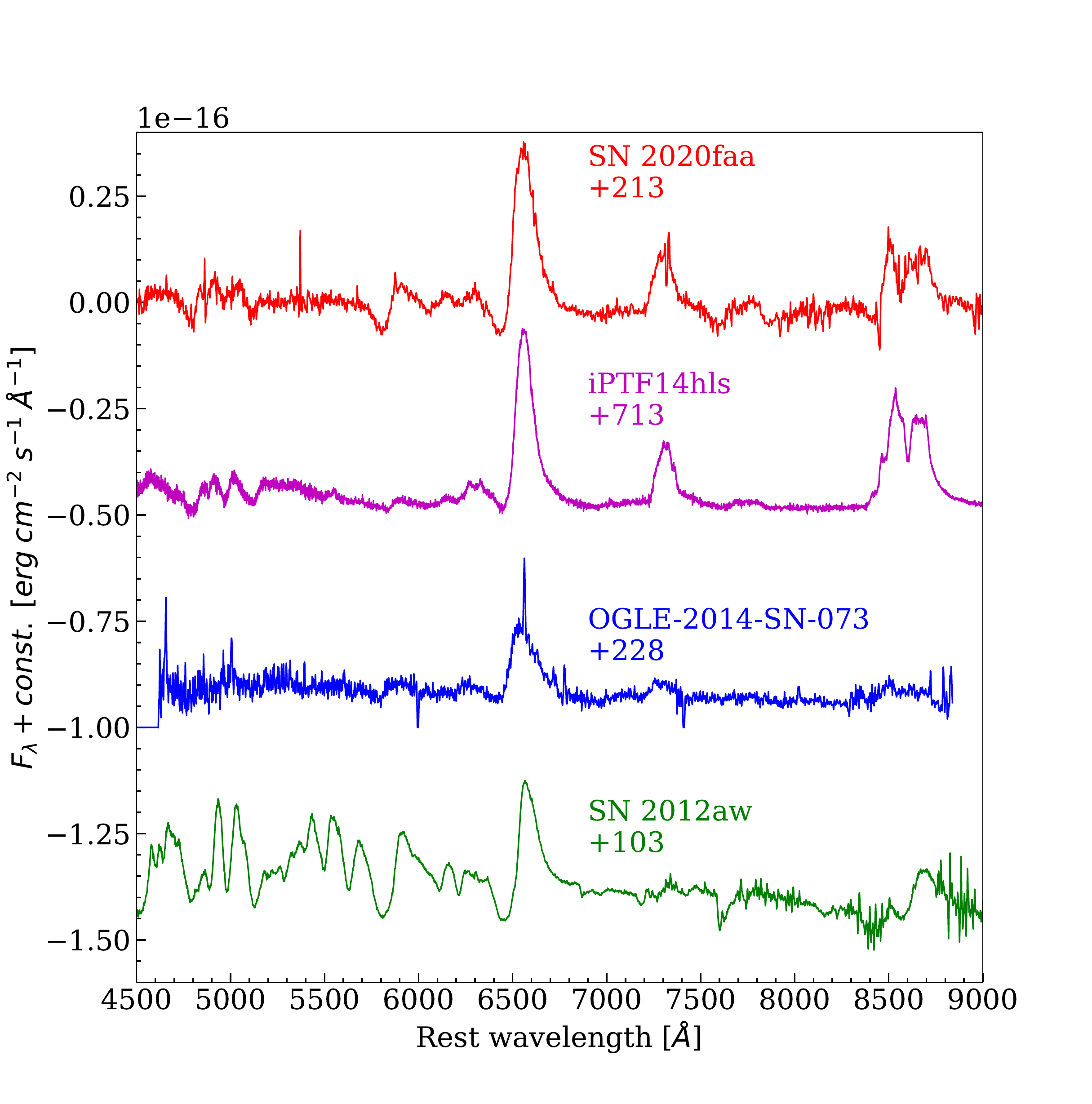}
        \caption{Spectrum of SN~2020faa at phase 213 d compared with spectra of other SNe in our sample. Spectra are taken at different phases for different SNe but are similar to each other, indicating different evolutionary timescales. Numbers near the spectra are days from explosion. All spectra are matched in intensity with respect to the H$\alpha$ emission and arbitrarily scaled for better visualisation.}
        \label{fig:spec_confronto_new}
    \end{figure}

In Fig.~\ref{fig:spec_confronto_new}, we plot the spectrum of SN~2020faa at phase 213~d and we compare it with the spectra of OGLE-2014-SN-073, iPTF14hls, and SN~2012aw (spectra from \cite{terreran_2017,arcavi_iPTF14hls_2017,bose_2012aw_2013}, respectively). The spectra were selected for the best match regardless of the phase. 
In fact, while for OGLE-2014-SN-073 the phase is almost the same as SN~2020faa (228~d instead of 213) and SN~2012aw has a faster spectral evolution, arriving at a similar stage of the spectral evolution after only 103 d. On the other hand, iPTF14hls is extremely slower and reaches similar features at 713~d. 

At this phase, SN~2020faa has clear P-Cygni profiles. The main emissions come from Balmer lines, with the addition of Fe~II $\lambda5169$, Na~I~D $\lambda\lambda5890,5896$, [Ca ~I]$\lambda \lambda7293,7326$, O~I $\lambda 7774$, and the blend of O~I $\lambda 8448$ with the Ca~II~NIR triplet $\lambda\lambda\lambda8498,8542,8662$.
OGLE-2014-SN-073 shows the same main emission lines with a P-Cygni profile that is slightly shallower than for SN~2020faa, but the Ca is significantly less strong. Moreover, it shows emission from [O~I]$\lambda\lambda6300,6364$, which SN~2020faa lacks.
Also, iPTF14hls shows the same emission lines of SN~2020faa, however, the Fe~II has a shallower P-Cygni profile, similarly to that of Ca~NIR triplet, and stronger Na~I~D.
Then, we see SN~2012aw has deep P-Cygni profiles in H, Na~I~D, and Fe~II, and stronger metal lines than SN~2020faa. A higher flux from the H lines could be due to a higher ejecta mass, but in general, the strongest lines are present in all SNe, while the weakest are seen only in SN~2012aw.

Overall, both iPTF14hls and OGLE-2014-SN-073 share many similarities with SN~2020faa. Spectrally speaking, iPTF14hls is more similar in terms of chemical composition and kinematics, but the time evolution is much slower. On the other hand, SN~2020faa and OGLE-2014-SN-073 have some differences in the spectra but the evolutionary phase is almost the same. Finally, we also recall that the light curve of OGLE-2014-SN-073 is the most similar to the one of SN~2020faa among the ones in our sample.


\section{Light curve interpretation}\label{sec:nagy}

As we noted above, SN~2020faa shares many similarities with OGLE-2014-SN-073 and also iPTF14hls and, as for these other two SNe, there is no definite understanding of the explosion scenario and light curve powering mechanisms. To set the perimeter of the possible configurations, we start listing the distinctive observational constraints.

The first one is set by spectroscopy: a common feature of the three events is the very long persistence of H I lines with P-Cygni profile that extends for 200~d for SN~2020faa and OGLE-2014-SN-073 and up to 700~d for iPTF14hls. This line profile is usually attributed to resonance scattering in a fast-expanding envelope located above a hot photosphere that is sustained by energy deposition in the innermost layers (e.g., radioactive decay). 

Another important observed property is the bright luminosity at the break-out, indicating that the progenitor had a large radius at the time of explosion. Assuming the temperature derived from the SED fitting, we estimated $R_0 \simeq 2 \times 10^{15}$ cm, which is one order of magnitude larger than the limit for a star in hydrostatic equilibrium. This suggests the presence of an extended circumstellar envelope produced in a very recent mass outflow.

At the same time, the lack of narrow H$\alpha$ emission argues against the presence of a slow-moving, massive stellar wind and there are no immediate signatures of CSM/ejecta interaction. If this even occurs, it is confined below the photosphere, while the kinematic of the outer envelope is consistent with a homologous expansion powered by the explosive shock wave.

After the break-out, the light curve shows a slow linear decline lasting $\sim60$~d and then a very broad peak that is most likely a signature of a long diffusion time and hence a large ejecta mass. Eventually, both SN~2020faa and OGLE-2014-SN-073 show a rapid drop in luminosity at 150-200 d  and then settle on a slow declining tail, which is better defined in SN~2020faa. This feature is not seen in iPTF14hls, which instead shows an irregular long-term decline. The tail of SN~2020faa is reminiscent of a radioactivity-powered light curve, although it is somewhat slower than the expectation from the decay time of $^{56}$Co.

With the aim of translating qualitative into quantitative constraints and to help explore the viable alternative models, we use as a guiding tool, the semi-analytical model originally introduced by \cite{arnett1980}, which has been found to provide a good representation of the light curve of H-rich core-collapse SNe. 
The model assumes a homologous expanding and spherically symmetric SN ejecta with a uniform density profile. Radiation transport is treated by the diffusion approximation and the effect of recombination, causing the rapid drop in the effective opacity in the envelope, is accounted for with a simple scheme described in \cite{arnett1989}.

It is well known that this model cannot fit the early phases, when the light curve is dominated by the SBO cooling phase. This feature is more evident in some SNe~IIb, whose light curves show a double peak. To deal with this type of events, \cite{nagy2016} proposed a two components structure for the SN ejecta. The idea is that the first peak of the light curve is dominated by the adiabatic cooling of the shock-heated, H-rich envelope, and the second peak is powered by radioactive decay deposited in the denser inner region. As stressed by \cite{nagy2016}, this progenitor structure may be explained by assuming an extended, low-mass envelope that is ejected just before the explosion. This two-component ejecta configuration was also tested with full hydrodynamic numerical simulations (e.g. references in \citealt{nagy2016}), showing that the semi-analytical approximation provides consistent results and is therefore suitable for an exploratory analysis aimed to set preliminary constraints of the parameter space.

As a further improvement, in their implementation, \cite{nagy2016} allowed for an alternative to radioactive decay energy input, considering the deposition in the ejecta of the rotational energy of a newborn magnetar. This latter option is often considered to explain the light curve of bright SNe, especially when the assumption of radioactive powering would require an implausibly large Ni mass. The magnetar engine has the advantage that with some freedom in the assumption for the spin decline rate, it is possible to accommodate for different slopes of the light curve tails. 

A full description of the parameters required by the modelling can be found in \cite{nagy2014} and \cite{nagy2016}.
For the specific case of SN~2020faa, the model parameters that we tried to constrain are: the initial radius, $R_0$, the ejected mass, $M_0$; the initial kinetic energy, $E_{kin}(0)$ and the initial thermal energy, $E_{th}(0)$, for both the shell and core components.  Also, we want to estimate the mass of radioactive material, $M_{Ni}$, or, alternatively, the initial rotational energy of the magnetar, $E_p$, and its spin-down rate, $t_{spin}$. 
Instead, for the additional parameters, we referred to the assumptions of \cite{nagy2016}, namely, $T_{rec} = 5500$K for the recombination temperature and $\kappa=0.3,0.4\,\mathrm{cm^2\,g^{-1}}$ for the opacity of the core and shell component, respectively.

\begin{figure}
    \centering
    \includegraphics[width=\columnwidth]{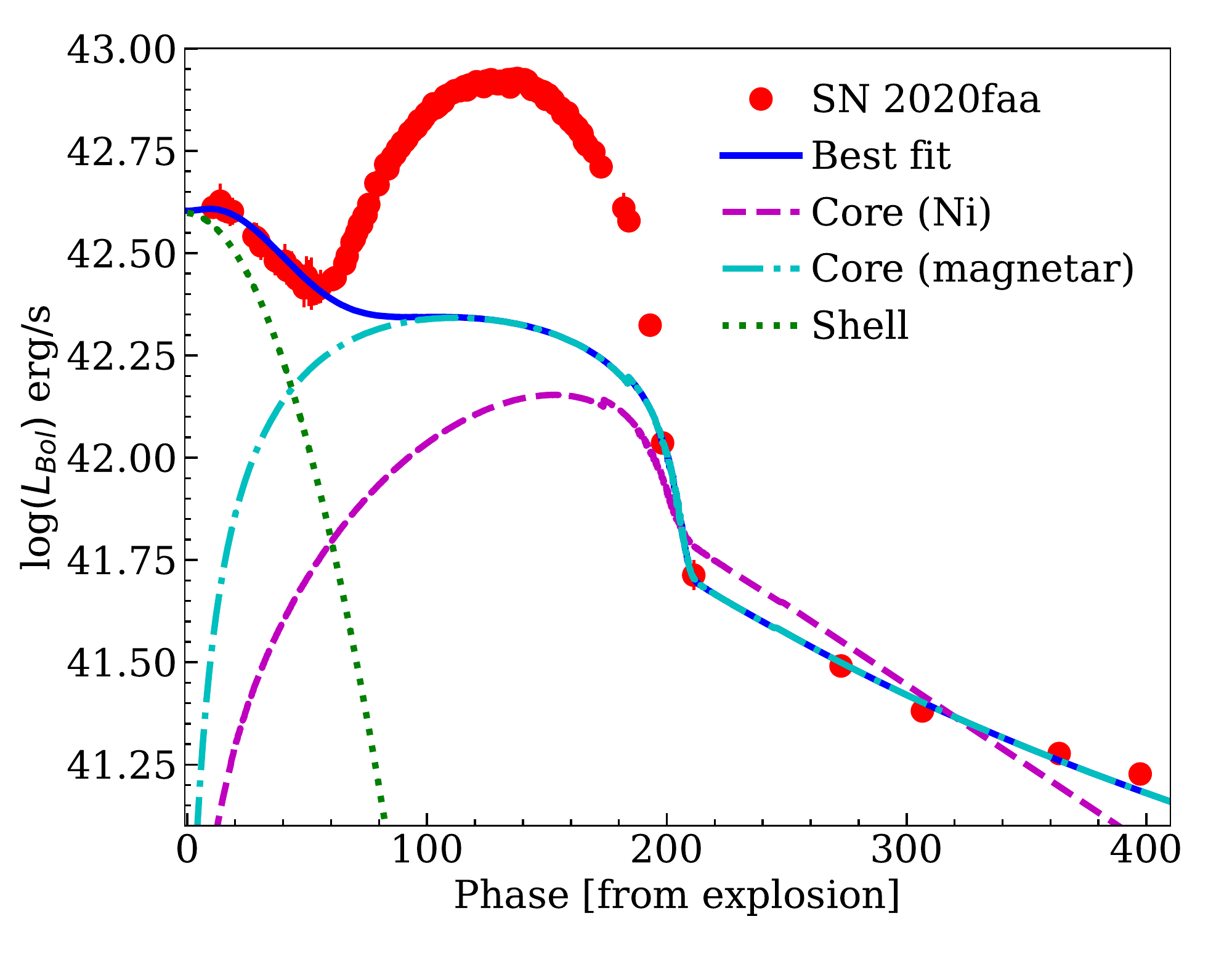}
    \caption{Best fit (solid blue line) of the bolometric luminosity (red dots) with the code LC2 by \cite{nagy2014} with a shell component (dotted green line) and a core+magnetar component (dash-dotted cyan line). The model does not provide a satisfactory power explanation for the peak, which requires an additional source (see text). An alternative fit with a core+Ni component, which yields a worse fit than the magnetar, is also plotted with a dashed magenta line.}
    \label{fig:nagy}
\end{figure}

We performed the fit by varying the free parameters on a grid and generating a different model for every variation. We then calculate the difference between the model and the observations, compute the $\chi^2$, and take the model corresponding to the minimum $\chi^2$ as a best fit, the result of which is shown in Fig.~\ref{fig:nagy}. 
A first exploratory analysis shows that, for a broad range of parameters allowing us to fit the break-out cooling phase, the shell contribution to the luminosity is dominant in the first $\sim$60 d, while in the second peak and, even more, in the tail, it is negligible. This can clearly be seen in Fig.~\ref{fig:nagy}, where the shell component contributes only during the initial decline. On the other hand, there seems to be no combination of parameters for the core component that allows fitting both the second maximum and the late-time tail. The problem is that the luminosity contrast between the second peak and the tail is far too bright for a regularly declining central energy input.

In fact, fitting the tail luminosity requires $M_{Ni}=0.28\pm 0.02\,\mathrm{M_\sun}$  or a powerful magnetar, $E_p = 1.5^{+0.5}_{-0.2} \times 10^{50}$ erg,  $t_{spin}=15\pm1$ d. As seen in  Fig.~\ref{fig:nagy}, the observed slope of the tail is shallower than the prediction from radioactive input, whereas tuning the magnetar spin-down rate allows for a perfect fit. We caution that the energy and spin-down timescale are degenerate parameters, therefore, there is a range of possible solutions to fit the slope. The values reported here are those corresponding to the minimum $\chi^2$. Also,
 we notice that recent studies on SNe Ia highlighted that ``Arnett-like'' models may overestimate $M_{Ni}$ when calculated at the peak \citep{khatami2019}. Moreover, it has been shown that the overestimate can go up to a factor of 2 for stripped-envelope SNe \citep{Afsariardchi2021}; however, this is still not enough to justify the amount of $^{56}$Ni required to fit the peak of SN~2020faa. Powering the second peak with the adopted model would require $M_{Ni}$ or $E_p$ values that would be one order of magnitude larger than the numbers above. This would argue against a standard neutrino-driven core-collapse explosion, for which the models indicate an upper limit of $M_{Ni}\sim0.3\,\mathrm{M_\sun}$ \citep{ni_2017}. Nevertheless, pair-instability explosions may be able to provide the missing few $\mathrm{M_\sun}$ of radioactive material \citep{kasen_pair_instability2011}. Alternatively, the required energy may be available from the spin-down of a newly born magnetar. In fact, when a magnetar powers the light curve,  a peak luminosity $\sim 10^{43}$ erg can be achieved with a magnetic field of $10^{14}\,\rm{G}$ and a spin-down of 2 ms \citep{kasen_magnetar_2010}. These parameters are consistent with a spin-down timescale of $\sim19$~d following \cite{cosimo_magnetar_2013} (their equation D7), a number very similar to what we obtain in our fit.
Moreover, \cite{Dessart2018} showed that a magnetar with $E_p=4 \times 10^{50}$ erg can provide the energy to power the long-term light curve of iPTF14hls. However, \cite{sollerman_iptf_2019} argued that the very late decline of the optical light curve of the transient is difficult to reconcile with this powerful central engine. This is even more critical in SN~2020faa because of the high luminosity contrast between the second peak and the tail.  

Hereafter, we used the light curve tail to constrain a radioactive or magnetar-powered model and tune the model parameters with the aim to match the observed light curve as closely as possible.
The result of this exercise is shown in Fig.~\ref{fig:nagy}. The fit of the early decline as SBO cooling indicates $M_{shell} = 2.4^{+0.5}_{-0.4}\,\mathrm{M}_\sun$, $E_{kin}(shell) = 0.9^{+0.5}_{-0.3} \times 10^{51}$ erg, with an initial radius $R_0 = 10^{14}$ cm and $E_{th} = 2.5 \times 10^{50}$ erg.  Instead, the phase of the drop from the second peak establishes the diffusion time in the core. We found that the luminosity at 100-200 d is maximised assuming $M_{core} = 21.5^{+1.4}_{-0.7}\,\mathrm{M}_\sun$, $E_{kin}(core) = 3.9^{+0.1}_{-0.4} \times 10^{51}$ erg ($R_0(core)= 2 \times 10^{13}$ cm and $E_{th}(core) = 1.0 \times 10^{50}$).
From these parameters, we can also derive an estimate of the expansion velocity (see \cite{nagy2016}, their Eq. 10), which varies from $v\sim9000\;\mbox{km\,s}^{-1}$ at the beginning of the evolution to $v\sim3000\;\mbox{km\,s}^{-1}$ at the end, consistently with what we measure from our spectra.
These numbers are summed up to give a progenitor mass at the time of explosion of at least $25\,\mathrm{M}_\sun$ (including the compact remnant) that appears consistent with the initial mass suggested by the environment study in Sec.~\ref{sec:host}. While these numbers do appear plausible, Fig.~\ref{fig:nagy} shows that even a two-component model of this sort is unable to provide a consistent fit of the whole light curve. In particular, we are not able to account for the high luminosity at the second peak. Subtracting the model from the observed light curve,  in $L_+ = 4.7 \times 10^{49}$ erg, we quantified the integrated luminosity excess. This energy needs to be extracted from some additional, short-lasting source that is located and thermalised below the photosphere to allow for the persistence of the P-Cygni broad absorption.

\section{Discussion}
\label{sec:interpr}

To address the peculiarities of the light curve of SN~2020faa, a first interesting comparison is with OGLE-2014-SN-073, whose bright, slow light curve was discussed to be consistent with a pair-instability explosion of a very massive star \citep{terreran_2017}. We notice that tight constraints of the epoch of explosion are not available for this transient and, therefore, we can make the similarity with SN~2020faa even more compelling by matching the time of the main peak of the two SNe. \cite{terreran_2017} found a reasonable match of the light curve of OGLE-2014-SN-073 with the pair instability model of \cite{Dessart2013}, in particular, the one labelled B190: a blue supergiant of initial mass $190\, \mathrm{M_{\sun}}$. Compared with SN~2020faa, the light curve of the B190 model is substantially broader and, more significantly, it has a much brighter tail. In fact, the model predicts $3\,\mathrm{M_{\sun}}$ of $^{56}$Ni in the ejecta, which allows us to explain the high luminosity at the peak, but this is in conflict with the constraint from the light curve tail of SN~2020faa (allowing for a Ni mass that is one order of magnitude lower). As already stated in Sec.~\ref{sec:nagy}, matching the $^{56}$Ni on the tail creates the need to account for missing energy in the peak.

A similar problem is present in the already mentioned magnetar engine proposed by \cite{Dessart2018} to explain the light curve of iPTF14hls. Again, with this interpretation, it is impossible to fit (with the same magnetar parameters) the bright peak and the subsequent rapid luminosity decline of SN~2020faa, unless there is a sudden change in the magnetar energy deposition.

We also recall that \cite{moriya_ogle_2018} were able to fit the light curve of OGLE-2014-SN-073 with a fallback-accretion model,  deriving a progenitor mass of $\sim30-40\,\mathrm{M}_\sun$, and explosion energy of $\sim 4\times 10^{51}$ erg. However, applying the same model to SN~2020faa is 
more challenging because of the complex luminosity evolution that would probably require an ad-hoc variation of the accretion rates.
Two alternative interpretations proposed for iPTF14hls appear more appealing to the specific case of SN~2020faa. Thus, they deserve a closer look and we discuss them in the following paragraphs.

\subsection{Hidden interaction with an inner disc}
For iPTF14hls, \cite{andrews_iptf_2018} proposed that additional input energy may come from the conversion of kinetic energy after interaction of the denser ejecta with a dense circumstellar disc or torus (see also \cite{McDowell2018,Nagao2020}).
\cite{smith2015} proposed the presence of a disc-like structure with a radius of about $1.5\times 10^{14}$ cm surrounding the progenitor. In this case, there would be narrow lines in emission in the very early days. The narrow H$\alpha$ that we see in the spectrum at +20~d could corroborate this hypothesis also for SN~2020faa. Given the close proximity of the disc, strong interaction is bound to happen soon and slow down the shock on the equatorial plane, but not in the polar region, where the ejecta will keep expanding freely, enshrouding the disc. Thus, interaction cannot be directly seen but it keeps heating up the inner part of the ejecta, giving a large contribution to the luminosity. Eventually, the photosphere will recede enough to reveal the interaction, which at this point will show up with lines of intermediate velocity due to the shock acceleration and possibly with double peaks if the interaction is still ongoing \citep{Smith_interaction_handbook_2017}.

This configuration was supported by the observation of double-peaked $H\alpha$ emission in very late spectra of iPTF14hls (at 1153~d). The idea is that the disc had a relatively small outer radius and was rapidly overrun by the outer ejecta, with strong interactions occurring when the disc was shocked by the inner, higher-density ejecta. When this happened, energy deposition occurred below the photosphere without further evidence if not for the high luminosity. A scenario with interaction taking place during the peak would also explain the increase in temperature we note in Fig.~\ref{fig:TeR}, being due to the extra heating supplied by the shock.

Considering this mechanism for SN~2020faa, we argue that to power the observed luminosity excess of the second peak, we need to use a fraction of the ejecta kinetic energy $f = L_+ / (\alpha\, E_{kin})$, where $\alpha<1$ accounts for the fact that only a fraction of the  kinetic energy may be converted into radiation. Assuming a typical value, $\alpha\sim0.25$ \citep{moriya_conversion_2006gy_2013}, we obtain $f=0.04$. This corresponds, for a homogeneous distribution, to the kinetic energy stored in $\sim 1 \,\mathrm{M_{\sun}}$ of the ejecta. In a fully inelastic collision, the shock will continue until the mass of the sweeping-up ejecta equals the mass of the disc that is therefore also $\sim 1 \mathrm{M_\sun}$ \citep{moriya_conversion_2006gy_2013}. This number is similar to that proposed by \cite{andrews_iptf_2018} for iPTF14hls. These authors also suggest that the disc extension was $10^{14.5} - 10^{15}$ cm, based on kinematic arguments. Such dimensions will work also for SN~2020faa, allowing for the onset of strong interaction a few weeks after explosion, followed by exhaustion a few months later. 


Nevertheless, we stress that the emission lines in the spectra of SN~2020faa do not show the double profiles that  have been claimed (in iPTF14hls) to support the presence of a disc structure in the CSM, nor do we notice intermediate-velocity components; our only favourable clue is the narrow H$\alpha$ at phase +20~d. While this could be attributed to an unfavourable orientation of the system, with the disc plane perpendicular to the line of sight, it is not easy to identify an evolutionary path that can produce the complex system configuration. Perhaps, the disc could be the result of a merging event that produced the progenitor of SN~2020faa, as proposed by \cite{smith_etacar_2018} to explain the formation of $\eta$ Carinae, but in this case, we would expect a more complex structure such as the Homunculus Nebula bipolar emission; however, we have no evidence of this in our case.

\subsection{Delayed, choked jet } 
A promising alternative for an impulsive inner engine calls for the onset of delayed jets. In this scenario, relativistic bipolar outflows are launched weeks or months after the explosion following mass accretion on the compact remnant, a neutron star or a black hole \citep{chugai_iptf_2018,Soker2018,akashi_jet_2022}. The jets are chocked in the massive ejecta and their energy is almost completely converted into radiation.  

In the model of \cite{chugai_iptf_2018}, the entire luminosity of iPTF14hls, after the first few weeks, can be attributed to the relativistic bipolar outflow produced by disc accretion onto a black hole, where the progenitor has a large ejecta mass, $M_{ej} = 30\,\mathrm{M}_\sun$, and kinetic energy of $E_{kin} = 8 \times 10^{51}$ erg. Also, the slow early luminosity decline is determined by the long diffusion time of the radiation of explosion energy. In this scenario, the SN ejecta structure is similar to our reference model for SN~2020faa.
However, there are two crucial observations that distinguish SN~2020faa:
i) the epoch of explosion of SN~2020faa is constrained to $\sim10$~d before discovery, while the models for iPTF14hls, in the lack of a tighter constraint for this transient, were allowed to explode 100-400~d before discovery and ii) the slow light curve tail of SN~2020faa, reminiscent of radioactive or magnetar slow-declining energy inputs. iPTF14hls may show something similar only at a much later time and with a steeper decline \citep{sollerman_iptf_2019}.

The constraints on the time of explosion and the initial high luminosity require a large initial radius ($10^{14}$ cm), likely that of a dense shell of material ejected shortly before explosion. On the other hand, the linear light curve tail seems to support the presence of a magnetar, rather than a black hole, as compact remnant. 
Overall, the system configuration may be very similar to that studied by \cite{kaplan2020} where a central object, either a neutron star or a black hole, accretes fallback material and launches two short-lived opposite jets weeks to months after the explosion. The jets interact with the ejecta and provide additional luminosity within the above time interval. The mechanism is very efficient, requiring, in the contest of iPTF14hls,  only 2\% of the SN explosion energy and a shock involving a small fraction of the ejecta mass to produce a luminosity of $\sim 10^{49} \,\mathrm{erg s^{-1}}$. This can also be applied to SN~2020faa.

\cite{Soker2022} claimed that jets in core-collapse SN explosions are ubiquitous (additional references therein).
Depending on progenitor configuration, energetics, and jet launching time, different types of transients can be produced. In addition, a jet will introduce asymmetries in the ejecta that may explain peculiar features also depending on the viewing angle.

\section{Conclusions}
\label{sec:conclusion}
SN~2020faa is a peculiar transient sharing many similarities with OGLE-2014-SN-073 and iPTF14hls, but also with unique characteristics that may help to shed some light on the origin of these unusual transients.

The long photospheric phase and broad light curve offer strong evidence that SN~2020faa at the time of explosion was a massive star of at least $M> 30 M_\sun$. A similar, high initial mass is also supported by the location of the transient in a young starburst region.

The bright initial luminosity and slow decline in the first 30 days indicate a very extended envelope ($10^{14}$ cm), likely due to a very recent outflow. The linear tail in the late light curve is indicative of a regularly declining energy injection by radioactive decay or, more likely, a fast-rotating magnetar.

The broad peak between 100-200~d cannot be explained by a central engine alone and requires a time-limited energy supplement, most likely coming from an inner or hidden shock. This can be generated by interaction of the denser ejecta with the surrounding CSM, if the latter is positioned very close to the progenitor and arranged in a disc or torus so that it is rapidly engulfed by the ejecta; thus it would not produce persistent narrow features in the spectra, but only a contribution to the overall luminosity. 
Another possibility is that delayed, relativistic polar outflows develop after disc mass accretion onto the compact remnant and shock the ejecta, so that their kinetic energy is converted into radiation. The delayed jet interpretation is particularly appealing because it can explain many other types of peculiar transients, depending on the dynamics of the jet and the viewing angle. 
In both interpretations, the shock region remains hidden deep into the ejecta and it is only manifested with the additional luminosity and we find they are equally likely to explain the features of SN~2020faa.

A disc or torus configuration interacting with the SN ejecta is interesting also in the context of a multimessenger scenario, since it is a suitable environment for particle acceleration and production of high-energy neutrinos \citep{fang2020}. Delayed jets, on the other hand, in spite of the fact that they provide adequate conditions for particle acceleration, seem unable to produce high-energy neutrinos \citep{guarini2023}.
We note that hidden interaction is a privileged location for the origin of high-energy neutrinos, even when this is not associated with the observation of high-energy photons. Transients such as SN~2020faa, with a long-lasting light curve, P-Cygni profiles, and slow spectral evolution, are also important for understanding the fate of very massive stars and their compact remnant. Therefore, keeping track of these peculiar SNe is fundamental for future multimessenger searches.


%

\begin{acknowledgements}
This paper is supported by fundings from MIUR, PRIN 2017 (grant 20179ZF5KS).
IS, EC, SB, MTB, NER, AP, and LT are partially supported by  the PRIN-INAF 2022 project "Shedding light on the nature of gap transients: from the observations to the models".
IS acknowledges the support of the doctoral grant funded by Istituto Nazionale di Astrofisica via the University of Padova and the Italian Ministry of Education, University and Research (MIUR).
AR acknowledges support from ANID BECAS/DOCTORADO NACIONAL 21202412.
GV acknowledges INAF for funding his PhD fellowship within the PhD School in Astronomy at the University of Padova.
We thank Daniel Perley for providing the spectrum taken with the Liverpool Telescope and the anonymous referee for their helpful comments.
This paper is based on observations collected at Copernico telescope (Asiago, Italy) of the INAF - Osservatorio Astronomico di Padova. 
Based on observations made with the Nordic Optical Telescope, owned in collaboration by the University of Turku and Aarhus University, and operated jointly by Aarhus University, the University of Turku and the University of Oslo, representing Denmark, Finland and Norway, the University of Iceland and Stockholm University at the Observatorio del Roque de los Muchachos, La Palma, Spain, of the Instituto de Astrofisica de Canarias. 
The Liverpool Telescope is operated on the island of La Palma by Liverpool John Moores University in the Spanish Observatorio del Roque de los Muchachos of the Instituto de Astrofisica de Canarias with financial support from the UK Science and Technology Facilities Council.
The Gran Telescopio CANARIAS (GTC) is a 10.4m telescope with a segmented primary mirror. It is located in one of the top astronomical sites in the Northern Hemisphere: the Observatorio del Roque de los Muchachos (ORM, La Palma, Canary Islands). The GTC is a Spanish initiative led by the Instituto de Astrofísica de Canarias (IAC). The project is actively supported by the Spanish Government and the Local Government from the Canary Islands through the European Funds for Regional Development (FEDER) provided by the European Union. The project also includes the participation of Mexico (Instituto de Astronomía de la Universidad Nacional Autónoma de México (IA-UNAM) and Instituto Nacional de Astrofísica, Óptica y Electrónica (INAOE), and the US University of Florida. 
We acknowledge the use of public data from the Swift data archive.
     
\end{acknowledgements}

\bibliographystyle{aa}
\bibliography{biblio}




%
%

\end{document}